\documentclass[letterpaper, 12pt]{article}[2000/05/19]
\usepackage[english]{babel}
\usepackage{amsfonts,amsmath,amssymb,amsthm,latexsym,amscd,mathrsfs}
\usepackage{ifthen,cite}
\usepackage[bookmarksnumbered=true]{hyperref}

\hypersetup{pdfpagetransition={Split}}

\newcommand{\evenhead}{Author \ name}
\newcommand{\oddhead}{Article \ name}
\newcommand{\theArticleName}{Article \ name}

\newcommand{\FirstPageHeading}[1]{\thispagestyle{empty}%
\noindent\raisebox{0pt}[0pt][0pt]{\makebox[\textwidth]{\protect\footnotesize \sf }}\par}

\newcommand{\ArticleName}[1]{\renewcommand{\theArticleName}{#1}\vspace{-2mm}\par\noindent {\LARGE\bf  #1\par}}
\newcommand{\Author}[1]{\vspace{5mm}\par\noindent {\Large  #1\par} \par\vspace{2mm}\par}
\newcommand{\Address}[1]{\vspace{2mm}\par\noindent {\it #1} \par}
\newcommand{\Email}[1]{\ifthenelse{\equal{#1}{}}{}{\par\noindent {\rm E-mail: }{\it  #1} \par}}
\newcommand{\EmailD}[1]{\ifthenelse{\equal{#1}{}}{}{\par\noindent {$\phantom{\dag}$~\rm E-mail: }{\it  #1} \par}}
\newcommand{\Abstract}[1]{\vspace{6mm}\par\noindent\hspace*{10mm}
\parbox{140mm}{\small {\bf Abstract.} #1}\par}
\newcommand{\Keywords}[1]{\vspace{3mm}\par\noindent\hspace*{10mm}
\parbox{140mm}{\small {\bf Key words:} \rm #1}\par}
\newcommand{\Classification}[1]{\vspace{3mm}\par\noindent\hspace*{10mm}
\parbox{140mm}{\small {\it 2010 Mathematics Subject Classification:} \rm #1}\vspace{3mm}\par}
\newcommand{\ShortArticleName}[1]{\renewcommand{\oddhead}{#1}}
\newcommand{\AuthorNameForHeading}[1]{\renewcommand{\evenhead}{#1}}

\long\def\@makecaption#1#2{
  \sbox\@tempboxa{\small \textbf{#1.}\ \ #2}%
  \ifdim \wd\@tempboxa >\hsize
    {\small \textbf{#1.}\ \ #2}\par \else
    \global \@minipagefalse
    \hb@xt@\hsize{\hfil\box\@tempboxa\hfil}%
  \fi \vskip\belowcaptionskip}

\def\numberwithin#1#2{\@ifundefined{c@#1}{\@nocounterr{#1}}{%
  \@ifundefined{c@#2}{\@nocnterr{#2}}{%
  \@addtoreset{#1}{#2}%
  \toks@\@xp\@xp\@xp{\csname the#1\endcsname}%
  \@xp\xdef\csname the#1\endcsname
    {\@xp\@nx\csname the#2\endcsname.\the\toks@}}}}
\def\E^#1{{\buildrel #1 \over\vee}}

{\theoremstyle{definition}

}

\usepackage{tikz}

\begin{document}

\FirstPageHeading{V.I. Gerasimenko, I.V. Gapyak}
\ShortArticleName{A brief survey}

\AuthorNameForHeading{V.I. Gerasimenko, I.V. Gapyak}

\textcolor{blue!50!black}{\ArticleName{The Boltzmann--Grad asymptotic behavior \\ of collisional dynamics: a brief survey}}

\Author{V.I. Gerasimenko$^*$\footnote{E-mail: \emph{gerasym@imath.kiev.ua}}
        \, and I.V. Gapyak$^{**}$\footnote{E-mail: \emph{gapjak@ukr.net}}}

\Address{* Institute of Mathematics of the NAS of Ukraine,\\
    3, Tereshchenkivs'ka Str.,\\
    01004, Kyiv, Ukraine}

\Address{** Taras Shevchenko National University of Kyiv,\\
    Department of Mechanics and Mathematics,\\
    2, Academician Glushkov Av.,\\
    03187, Kyiv, Ukraine}

\bigskip
\Abstract{
The article discusses some of the latest advances in the mathematical understanding of the nature of kinetic
equations that describe the collective behavior of many-particle systems with collisional dynamics.
}

\bigskip

\bigskip
\Classification{82C05, 82C40, 35Q20, 46N55, 47J35}

\makeatletter
\renewcommand{\@evenhead}{
\hspace*{-5pt}\raisebox{-15pt}[\headheight][0pt]{\vbox{\hbox to \textwidth {\thepage \hfil \evenhead}\vskip4pt \hrule}}}
\renewcommand{\@oddhead}{
\hspace*{-5pt}\raisebox{-15pt}[\headheight][0pt]{\vbox{\hbox to \textwidth {\oddhead \hfil \thepage}\vskip4pt\hrule}}}
\renewcommand{\@evenfoot}{}
\renewcommand{\@oddfoot}{}
\makeatother

\newpage
\vphantom{math}

\protect\textcolor{blue!50!black}{\tableofcontents}

\newpage

\bigskip
\textcolor{blue!50!black}{\section{Overview}}

The purpose of the paper is to review some of the achievements in the mathematical
understanding of kinetic equations for many-particle systems with collisional dynamics.

As known, on the basis of the description of many-particle systems is the notions of
the state and the observable. The functional of the mean value of observables defines
a duality between observables and states and as a consequence there exist two approaches
to the description of the evolution. Usually, the evolution of both finitely and infinitely
many classical particles is described within the framework of the state evolution by the
BBGKY (Bogolyubov--Born--Green--Kirkwood--Yvon) hierarchy for marginal (reduced) distribution
functions \cite{CGP97}. An equivalent approach to describing evolution consists in the
description of the evolution of observables by the dual BBGKY hierarchy for marginal (reduced)
observables \cite{BGer},\cite{BGer10}.

In certain situations, the collective behavior of many-particle systems can be adequately
described by the Boltzmann kinetic equation \cite{Sp91}-\cite{SR12}. The conventional
philosophy of the description of the kinetic evolution consists of the following. If the
initial state specified by a one-particle (marginal) distribution function, then the evolution
of the state can be effectively described by means of a one-particle (marginal) distribution
function governed in a suitable scaling limit by the nonlinear kinetic equation \cite{Sp80},\cite{SSD}.
The Boltzman--Grad asymptotic behavior \cite{GH} of a perturbative solution of the BBGKY
hierarchy for hard spheres with elastic collisions was constructed in \cite{C72}-\cite{PG90}
(in \cite{SR12},\cite{PSS} -- for short-range interaction potential).

In the survey, we consider an approach to the description of the kinetic evolution of infinitely
many hard spheres within the framework of the evolution of marginal observables. For this purpose,
the Boltzmann--Grad asymptotic behavior of a nonperturbative solution of the Cauchy problem of the
dual BBGKY hierarchy is constructed. \cite{G13},\cite{GG18}. In the case of initial states specified
by means of a one-particle distribution function the relations between the Boltzmann--Grad asymptotic
behavior of marginal observables and a solution of the Boltzmann kinetic equation is established.
In particular, the evolution of additive-type marginal observables is equivalent to a solution of
the Boltzmann equation and the evolution of nonadditive-type marginal observables is equivalent to
the property of the propagation in time of initial chaos for states.

One of the advantages of such an approach to the derivation of kinetic equations from the underlying
dynamics of hard spheres consists of an opportunity to construct the Boltzmann-type kinetic equation
with initial correlations, which makes it possible also to describe the propagation of the initial
correlations in the Boltzmann--Grad approximation \cite{GG18}.

Moreover, using the suggested approach, we can derive the non-Markovian Enskog kinetic equation \cite{GG}
and construct the marginal functionals of states, describing the creation and the propagation of
correlations of particles with hard sphere collisions in terms of a one-particle distribution function
that is governed by the stated Enskog equation.

The established results we extend on systems of hard spheres with inelastic collisions. In particular,
it was established that in a one-dimensional space the kinetic evolution of a large system of hard rods
with inelastic collisions is governed by the certain generalization of the known Boltzmann equation for
a one-dimensional granular gas \cite{BG}.

\bigskip
\textcolor{blue!50!black}{\section{Preliminaries: dynamics of finitely many hard spheres}}

\vskip-5mm
\textcolor{blue!50!black}{\subsection{Semigroups of operators of a hard sphere system}}
We consider a system of identical particles of a unit mass interacting as hard spheres with
a diameter of $\sigma>0$. Every particle is characterized by its phase coordinates
$(q_{i},p_{i})\equiv x_{i}\in\mathbb{R}^{3}\times\mathbb{R}^{3},\, i\geq1.$ For configurations of
such a system the following inequalities are satisfied: $|q_i-q_j|\geq\sigma,$ $i\neq j\geq1$,
i.e. the set $\mathbb{W}_n\equiv\big\{(q_1,\ldots,q_n)\in\mathbb{R}^{3n}\big||q_i-q_j|<\sigma$ for
at least one pair $(i,j):\,i\neq j\in(1,\ldots,n)\big\}$, $n>1$, is the set of forbidden configurations.

Let $C_n\equiv C(\mathbb{R}^{3n}\times(\mathbb{R}^{3n}\setminus \mathbb{W}_n))$ be the space of bounded
continuous functions on $\mathbb{R}^{3n}\times(\mathbb{R}^{3n}\setminus \mathbb{W}_n)$ that are symmetric
with respect to permutations of the arguments $x_1,\ldots,x_n$, equal to zero on the set of forbidden
configurations $\mathbb{\mathbb{W}}_n$ and equipped with the norm:
$\|b_n\|=\sup_{x_{1},\ldots,x_{n}}|b_n(x_{1},\ldots,x_{n})|$.

To describe dynamics of $n$ hard spheres we introduce the one-parameter mapping $S_{n}(t)$ on the space
$C_n\equiv C(\mathbb{R}^{3n}\times(\mathbb{R}^{3n}\setminus \mathbb{W}_n))$ by means of the phase
trajectories of a hard sphere system, which are defined almost everywhere on the phase space
$\mathbb{R}^{3n}\times(\mathbb{R}^{3n}\setminus \mathbb{W}_n)$, namely, beyond of the set
$\mathbb{M}_{n}^0$ of the zero Lebesgue measure, as follows
\begin{eqnarray} \label{Sspher}
  &&\hskip-5mm(S_{n}(t)b_{n})(x_{1},\ldots,x_{n})\equiv S_{n}(t,1,\ldots,n)b_{n}(x_{1},\ldots,x_{n})\doteq\\
  &&\hskip-5mm\begin{cases}
         b_{n}(X_{1}(t,x_{1},\ldots,x_{n}),\ldots,X_{n}(t,x_{1},\ldots,x_{n})),\\
         \hskip+45mm\mathrm{if}\,(x_{1},\ldots,x_{n})\in(\mathbb{R}^{3n}\times(\mathbb{R}^{3n}\setminus\mathbb{W}_{n})),\\
         0, \hskip+42mm \mathrm{if}\,(q_{1},\ldots,q_{n})\in\mathbb{W}_{n},
                    \end{cases}\nonumber
\end{eqnarray}
where for $t\in\mathbb{R}$ the function $\texttt{X}_{i}(t)$ is a phase trajectory of $ith$ particle
constructed in \cite{CGP97} and the set $\mathbb{M}_{n}^0$ consists of the phase space points which
are specified such initial data that during the evolution generate multiple collisions, i.e. collisions
of more than two particles, more than one two-particle collision at the same instant and infinite number
of collisions within a finite time interval \cite{CGP97},\cite{A}.

On the space $C_n$ one-parameter mapping (\ref{Sspher}) is an isometric $\ast$-weak continuous
group of operators, i.e. it is a $C_{0}^{\ast}$-group.

We introduce the notion of the $nth$-order cumulant of groups of operators (\ref{Sspher})
\begin{eqnarray}\label{cumulan}
    &&\mathfrak{A}_{n}(t,X)\doteq\sum\limits_{\mathrm{P}:\,X={\bigcup}_i X_i}
       (-1)^{\mathrm{|P|}-1}({\mathrm{|P|}-1})!\prod_{X_i\subset \mathrm{P}}S_{|X_i|}(t,X_i),
\end{eqnarray}
where ${\sum}_\mathrm{P}$ is the sum over all possible partitions $\mathrm{P}$ of the set
$X\equiv(1,\ldots,n)$ into $|\mathrm{P}|$ nonempty mutually disjoint subsets  $X_i\subset X$.

Let us indicate some properties of cumulants (\ref{cumulan}). If $n=1$, on the domain of the
definition $b_{1}\in\mathcal{D}\subset C_1$ in the sense of the $\ast$-weak convergence
of the space $C_1$ a generator of the first-order cumulant is given by the operator
\begin{eqnarray*}\label{Lni}
    &&\mathrm{w^{\ast}-}\lim\limits_{t\rightarrow 0}\frac{1}{t}(\mathfrak{A}_{1}(t,1)-I)b_{1}(x_1)
       =\mathcal{L}(1)b_{1}(x_1)\doteq \langle p_{1},\frac{\partial}{\partial q_{1}}\rangle b_{1}(x_1),
\end{eqnarray*}
where the symbol $\langle \cdot,\cdot \rangle$ means a scalar product.

In the case of $n=2$, if $t>0$, for $b_{2}\in\mathcal{D}\subset C_2$ in the
sense of a $\ast$-weak convergence of the space $C_2$ the following equality holds
\begin{eqnarray*}\label{Li}
   &&\hskip-7mm\mathrm{w^{\ast}-}\lim\limits_{t\rightarrow 0}\frac{1}{t}\,\mathfrak{A}_{2}(t,1,2)b_{2}(x_1,x_2)
      =\mathcal{L}_{\mathrm{int}}(1,2)b_{2}(x_1,x_2)\doteq \\
    &&\hskip-7mm\sigma^2\int_{\mathbb{S}_+^2}d\eta\langle\eta,(p_{1}-p_{2})\rangle
      \big(b_2(q_{1},p_{1}^*,q_{2},p_{2}^*)-b_2(x_1,x_2)\big)\delta(q_{1}-q_{2}+\sigma\eta),\nonumber
\end{eqnarray*}
where $\delta$ is the Dirac measure,
$\mathbb{S}_{+}^{2}\doteq\{\eta\in\mathbb{R}^{3}\big|\left|\eta\right|=1\langle\eta,(p_1-p_{2})\rangle>0\}$
and the momenta $p_{1}^*,p_{2}^*$ are determined by the equalities
\begin{eqnarray}\label{momenta}
     &&p_i^*\doteq p_i-\eta\left\langle\eta,\left(p_i-p_{j}\right)\right\rangle, \\
     &&p_{j}^*\doteq p_{j}+\eta\left\langle\eta,\left(p_i-p_{j}\right)\right\rangle \nonumber.
\end{eqnarray}
If $t<0$, the operator $\mathcal{L}_{\mathrm{int}}(1,2)$ is defined by the corresponding expression \cite{CGP97}.

In the case of $n>2$ as a consequence of the fact that for a hard sphere system group (\ref{Sspher})
is defined almost everywhere on the phase space $\mathbb{R}^{3n}\times(\mathbb{R}^{3n}\setminus \mathbb{W}_n)$,
i.e. there are no collisions of more than two particles at every instant, the equality holds
\begin{eqnarray*}
   &&\mathrm{w^{\ast}-}\lim\limits_{t\rightarrow 0}\frac{1}{t}\,\mathfrak{A}_{n}(t,1,\ldots,n)b_{n}(x_1,\ldots,x_n)=0.
\end{eqnarray*}

For the group of evolution operators (\ref{Sspher}) the Duhamel equation holds
\begin{eqnarray*}\label{DuamN}
    &&S_n(t,1,\ldots,n)b_n=\\
    &&\prod\limits_{i=1}^nS_1(t,i)b_n+\int\limits_0^td\tau \prod\limits_{i=1}^{n}S_{1}(t-\tau,i)
       \sum\limits_{j_1<j_2=1}^{n}\mathcal{L}_{\mathrm{int}}(j_1,j_2)S_{n}(\tau,1,\ldots,n)b_{n}=\nonumber\\
    &&\prod\limits_{i=1}^nS_1(t,i)b_n+\int\limits_0^td\tau S_{n} (t-\tau,1,\ldots,n)
       \sum\limits_{j_{1}<j_{2}=1}^{n}\mathcal{L}_{\mathrm{int}}(j_{1},j_{2})
       \prod\limits_{i=1}^{n}S_{1}(\tau,i)b_{n},\nonumber
\end{eqnarray*}
where the operator $\mathcal{L}_{\mathrm{int}}(j_{1},j_{2})$ is defined by the formula
\begin{eqnarray}\label{Lint}
     &&\mathcal{L}_{\mathrm{int}}(j_{1},j_{2})b_{n}\doteq
        \sigma^2\int_{\mathbb{S}_{+}^2}d\eta\langle\eta,(p_{j_{1}}-p_{j_{2}})\rangle
        \big(b_n(x_1,\ldots,p_{j_{1}}^\ast,q_{j_{1}},\ldots,\\
     &&\hskip+21mmp_{j_{2}}^\ast,q_{j_{2}},\ldots,x_n)-b_n(x_1,\ldots,x_n)\big)\delta(q_{j_{1}}-q_{j_{2}}+\sigma\eta).\nonumber
\end{eqnarray}

Thus, the infinitesimal generator $\mathcal{L}_{n}$ of the group of operators (\ref{Sspher})
has the structure
\begin{eqnarray}\label{L}
     &&\mathcal{L}_{n}b_{n}\doteq\sum\limits_{j=1}^{n}\mathcal{L}(j)b_{n}+
        \sum\limits_{j_{1}<j_{2}=1=1}^{n}\mathcal{L}_{\mathrm{int}}(j_{1},j_{2})b_{n},
\end{eqnarray}
where the operator
$\mathcal{L}(j)\doteq\langle p_j,\frac{\partial}{\partial q_j}\rangle$ defined on the set $C_{n,0}$
we had denoted by symbol $\mathcal{L}(j)$ .

Let $L^{1}_{n}\equiv L^{1}(\mathbb{R}^{3n}\times(\mathbb{R}^{3n}\setminus \mathbb{W}_n))$ be the
space of integrable functions that are symmetric with respect to permutations of the arguments
$x_1,\ldots,x_n$, equal to zero on the set of forbidden configurations $\mathbb{W}_n$ and equipped
with the norm: $\|f_n\|_{L^{1}(\mathbb{R}^{3n}\times\mathbb{R}^{3n})}=\int dx_1\ldots dx_n|f_n(x_1,\ldots,x_n)|$.
Hereafter the subspace of continuously differentiable functions with compact supports we will denote by
$L_{n,0}^1\subset L^1_n$ and the subspace of finite sequences of continuously differentiable functions
with compact supports let be $L_{0}^1\subset L^{1}_{\alpha}=\oplus^{\infty}_{n=0}\alpha^n L^{1}_{n}$,
where $\alpha>1$ is a real number.

On the space of integrable functions we define the adjoint operator $S_n^\ast(t)$ to operator (\ref{Sspher}).
The following equality takes place
\begin{eqnarray}\label{ad}
   &&S_n^\ast(t)=S_n(-t).
\end{eqnarray}
This group satisfies the Duhamel equation
\begin{eqnarray*}\label{DuamN_1}
    &&S_n^\ast(t,1,\ldots,n)=\prod\limits_{i=1}^{n}S_1^\ast(t,i)+\int\limits_0^t d\tau \prod\limits_{i=1}^{n}S_{1}^\ast(t-\tau,i)
       \sum\limits_{j_{1}<j_{2}=1}^{n}\mathcal{L}_{\mathrm{int}}^\ast(j_{1},j_{2})S_{n}^\ast(\tau,1,\ldots,n)),
\end{eqnarray*}
where the operator $\mathcal{L}_{\mathrm{int}}^\ast(j_{1},j_{2})$ is defined by the formula
\begin{eqnarray}\label{bLint}
     &&\hskip-9mm \mathcal{L}_{\mathrm{int}}^\ast(j_{1},j_{2})f_{n}
        \doteq\sigma^2\int_{\mathbb{S}_{+}^2}d\eta\langle\eta,(p_{j_{1}}-p_{j_{2}})\rangle
        f_n(x_1,\ldots,p_{j_{1}}^*,q_{j_{1}},\ldots,\\
     &&p_{j_{2}}^*,q_{j_{2}},\ldots,x_n)\delta(q_{j_{1}}-q_{j_{2}}+\sigma\eta)-
        f_n(x_1,\ldots,x_n)\delta(q_{j_{1}}-q_{j_{2}}-\sigma\eta)\big).\nonumber
\end{eqnarray}
In formula (\ref{bLint}) the notations similar to (\ref{Lint}) are used and symbol $\delta$
is a Dirac measure.

On the space $L_n^1$ one-parameter mapping defined by (\ref{ad}) is an isometric strong continuous
group of operators. Indeed, $\big\|S_n^\ast(t)\big\|=1$.

We note also the validity of the equality
\begin{eqnarray*}
    &&\prod\limits_{i=1}^{n}S_1^\ast(t,i)+\int\limits_0^t d\tau \prod\limits_{i=1}^{n}S_{1}^\ast(t-\tau,i)
       \sum\limits_{j_{1}<j_{2}=1}^{n}\mathcal{L}_{\mathrm{int}}^\ast(j_{1},j_{2})S_{n}^\ast(\tau,1,\ldots,n)=\nonumber\\
    &&=\prod\limits_{i=1}^{n}S_1^\ast(t,i)+\int\limits_0^td\tau S_{n}^\ast(t-\tau,1,\ldots,n)
       \sum\limits_{j_{1}<j_{2}=1}^{n}\mathcal{L}_{\mathrm{int}}^\ast(j_{1},j_{2})\prod\limits_{i=1}^{n}S_{1}^\ast(\tau,i),\nonumber
\end{eqnarray*}

Thus, the infinitesimal generator $\mathcal{L}_{n}^\ast$ of the group of operators $S_n^\ast(t)$ has the structure
\begin{eqnarray*}\label{Lstar}
    &&\mathcal{L}_{n}^\ast f_{n}\doteq\sum\limits_{j=1}^{n}\mathcal{L}^\ast(j)f_{n}+
        \sum\limits_{j_{1}<j_{2}=1}^{n}\mathcal{L}_{\mathrm{int}}^\ast(j_{1},j_{2})f_{n},
\end{eqnarray*}
where the operator $\mathcal{L}^{\ast}(j)\doteq-\langle p_j,\frac{\partial}{\partial q_j}\rangle$
defined on the set $L_{n,0}^1\subset L^1_n$ we had denoted by the symbol $\mathcal{L}^{\ast}(j)$.

\textcolor{blue!50!black}{\subsection{The functional for mean values of observables}}
It is known \cite{CGP97} that many-particle systems are described in terms of observables and states.
The functional of the mean value of the observables determines a duality between the observables and
the states, and, as a result, there are two equivalent approaches to the description of the evolution.

We will consider a hard sphere system of a non-fixed, i.e. arbitrary but finite average number of identical
particles (nonequilibrium grand canonical ensemble) in the space $\mathbb{R}^{3}$. In this case, the
evolution of the observables $A(t)=(A_0,A_{1}(t,x_1),\ldots,A_{n}(t,x_1,\ldots,x_n),\ldots)$ describes
by the Cauchy problem for the sequence of the weak formulation of the Liouville (pseudo-Liouville) equations
\begin{eqnarray}
  \label{H-N1}
     &&\frac{d}{d t}A(t)=\mathcal{L}A(t),\\
  \label{H-N12}
     &&A(t)|_{t=0}=A(0),
\end{eqnarray}
where the operator $\mathcal{L}=\oplus_{n=0}^\infty\mathcal{L}_n$ is defined by formula (\ref{L}).

Let $C_{\gamma}$ be the space of sequences $b=(b_0,b_1,\ldots,b_n,\ldots)$ of bounded continuous functions
$b_n\in C_n$ equipped with the norm: $\|b_n\|_{C_{\gamma}}=\max_{n\geq 0}\,\frac{\gamma^{n}}{n!}\,\|b_n\|_{C_n}$.

If $A(0)\in C_{\gamma}$, then a unique solution of the Cauchy problem \eqref{H-N1},\eqref{H-N12} is
determined by the formula
\begin{eqnarray*}\label{sH}
    &&A(t)=S(t)A(0),
\end{eqnarray*}
where for $t\in\mathbb{R}$ the group of operators $S(t)=\oplus_{n=0}^\infty S_n(t)$ is defined
by formula \eqref{Sspher}.

The average values of observables (mean values of observables) are determined by the positive
continuous linear functional on the space $C_{\gamma}$
\begin{eqnarray}\label{averageD}
     &&\hskip-5mm \langle A\rangle(t)=(A(t),D(0))\doteq(I,D(0))^{-1}\sum\limits_{n=0}^{\infty}\frac{1}{n!}
         \,\int_{(\mathbb{R}^{3}\times\mathbb{R}^{3})^{n}}dx_{1}\ldots dx_{n}\,A_{n}(t)\,D_{n}^0,
\end{eqnarray}
where $D(0)=(1,D_{1}^0,\ldots,D_{n}^0,\ldots)$ is a sequence of symmetric nonnegative functions
that describes the states of a system of a non-fixed number of hard spheres and the normalizing factor
$(I,D(0))={\sum\limits}_{n=0}^{\infty}\frac{1}{n!}\int_{(\mathbb{R}^{3}\times\mathbb{R}^{3})^{n}}dx_{1}\ldots dx_{n}D_{n}^0$
is a grand canonical partition function. For the sequences $D(0)\in L^{1}_{\alpha}$ and $A(t)\in C_{\gamma}$ average
value functional \eqref{averageD} exists and determines a duality between observables and states.

Taking into account the equality $(I,D(0))=(I,S^\ast(t)D(0))$, and, as a consequence,  of the validity
for functional \eqref{averageD} of the following equality
\begin{eqnarray*}
     &&(A(t),D(0))=(I,D(0))^{-1}\sum\limits_{n=0}^{\infty}\frac{1}{n!}
         \,\int_{(\mathbb{R}^{3}\times\mathbb{R}^{3})^{n}}dx_{1}\ldots dx_{n}\,
         S_{n}(t)A_{n}^0\,D_{n}^0=\\
     &&(I,S^\ast(t)D(0))^{-1}\sum\limits_{n=0}^{\infty}\frac{1}{n!}\,
         \int_{(\mathbb{R}^{3}\times\mathbb{R}^{3})^{n}}dx_{1}\ldots dx_{n}\,
         A_{n}^0\,S^\ast_n(t)D_{n}^0\equiv\\
     &&(I,D(t))^{-1}(A(0),D(t)),
\end{eqnarray*}
we can describe the evolution within the framework of the evolution of states. Indeed, the evolution
of all possible states, i.e. the sequence $D(t)=(1,D_{1}(t),\ldots, D_{n}(t),\ldots)\in L^{1}_{\alpha}$
of the distribution function $D_{n}(t),\, n\geq1$, is described by the Cauchy problem for a sequence of
the Liouville equations (the dual pseudo-Liouville equations) for states
\begin{eqnarray}
  \label{vonNeumannEqn}
     &&\frac{d}{d t}D(t)=\mathcal{L}^\ast D(t),\\
  \label{F-N12}
     &&D(t)|_{t=0}=D(0).
\end{eqnarray}
The generator $\mathcal{L}^\ast=\oplus^{\infty}_{n=0}\mathcal{L}^\ast_{n}$ of the dual Liouville equations
\eqref{vonNeumannEqn} is the adjoint operator to generator \eqref{L} of the Liouville equation \eqref{H-N1}
in the sense of functional \eqref{averageD}, and it is defined by formula \eqref{Lstar}.

A unique solution of the Cauchy problem of the dual Liouville equation \eqref{vonNeumannEqn},\eqref{F-N12}
is determined by the formula
\begin{eqnarray*}\label{rozv_fon-N}
    &&D(t)=S^\ast(t)D(0),
\end{eqnarray*}
where the one-parameter family of operators $S^\ast(t)=\oplus_{n=0}^{\infty}S^\ast_{n}(t)$,
is defined as above.

For a system of a finite average number of hard spheres there exists an equivalent possibility to
describe observables and states, namely, by sequences of marginal observables (so-called the reduced or
$s$-particle observables) $B(t)=(B_0,B_{1}(t,x_1),\ldots,B_{s}(t,x_1,\ldots,x_s),\ldots)$ and marginal
states (reduced or $s$-particle distribution functions) $F(0)=(1,F_{1}^0(x_1),\ldots,F_{s}^0(x_1,\ldots,x_s),\ldots)$
\cite{CGP97}. These sequences are correspondingly introduced instead of sequences of the observables $A(t)$
and the distribution functions $D(0)$, in such way that their mean value \eqref{averageD} does not change, i.e.
\begin{eqnarray}\label{avmar-1}
   &&\hskip-12mm\big\langle A\big\rangle(t)=(I,D(0))^{-1}(A(t),D(0))=(I,D(0))^{-1}\sum\limits_{n=0}^{\infty}\frac{1}{n!}
      \,\int_{(\mathbb{R}^{3}\times\mathbb{R}^{3})^{n}}dx_{1}\ldots dx_{n}\,A_{n}(t)\,D_{n}^0=\\
   &&\hskip+12mm \sum\limits_{s=0}^{\infty}\frac{1}{s!}\,
      \int_{(\mathbb{R}^{3}\times\mathbb{R}^{3})^{n}}dx_{1}\ldots dx_{s}\,
      B_{s}(t,x_1,\ldots,x_s)\,F_{s}^0(x_1,\ldots,x_s)=(B(t),F(0)),\nonumber
\end{eqnarray}
where $(I,D(0))$ is a normalizing factor defined above and $I\equiv(1,\ldots,1,\ldots)$.

Thus, the relations of marginal observables and observables are determined by the following formula:
\begin{eqnarray*}\label{mo}
      &&B_{s}(t,x_1,\ldots,x_s)\doteq\sum_{n=0}^s\,\frac{(-1)^n}{n!}\sum_{j_1\neq\ldots\neq j_{n}=1}^s
            A_{s-n}(t,(x_1,\ldots,x_s)\setminus(x_{j_1},\ldots,x_{j_{n}})), \quad s\geq 1,
\end{eqnarray*}
where $(x_1,\ldots,x_s)\setminus(x_{j_1},\ldots,x_{j_{n}})\equiv(x_1,\ldots,x_{j_1-1},x_{j_1+1},
\ldots,x_{j_{n}-1},x_{j_{n}+1},\ldots,x_s)$.
In terms of the distribution functions, the marginal distribution functions are defined as follows:
\begin{eqnarray*}\label{ms}
      &&F_{s}^0(x_1,\ldots,x_s)\doteq(I,D(0))^{-1} \sum\limits_{n=0}^{\infty}\frac{1}{n!}
          \int_{(\mathbb{R}^{3}\times\mathbb{R}^{3})^{n}}dx_{s+1}\ldots dx_{s+n}
          \,D_{s+n}^0(x_1,\ldots,x_{s+n}),\quad s\geq 1,
\end{eqnarray*}
where the normalizing factor $(I,D(0))$ is defined above.

We note that traditionally \cite{CGP97}, the evolution of many-particle systems is described within the
framework of the evolution of states by the BBGKY hierarchy for marginal distribution functions. An
equivalent approach to describing evolution is based on marginal observables governed by the dual BBGKY
hierarchy.

\textcolor{blue!50!black}{\subsection{The dual BBGKY hierarchy for hard spheres}}
If $t\geq0$, the evolution of marginal observables of a system of a non-fixed number of hard spheres
is described by the Cauchy problem of the weak formulation of the dual BBGKY hierarchy \cite{BGer}:
\begin{eqnarray}\label{dh}
   &&\hskip-9mm\frac{\partial}{\partial t}B_{s}(t,x_1,\ldots,x_s)=\big(\sum\limits_{j=1}^{s}\mathcal{L}(j)+
      \sum\limits_{j_1<j_{2}=1}^{s}\mathcal{L}_{\mathrm{int}}(j_1,j_{2})\big)B_{s}(t,x_1,\ldots,x_s)+\\
   &&\hskip-9mm\sum_{j_1\neq j_{2}=1}^s
      \mathcal{L}_{\mathrm{int}}(j_1,j_{2})B_{s-1}(t,x_1,\ldots,x_{j_1-1},x_{j_1+1},\ldots,x_s),\nonumber\\
      \nonumber\\
  \label{dhi}
  &&\hskip-9mm B_{s}(t,x_1,\ldots,x_s)_{\mid t=0}=B_{s}^{\epsilon,0}(x_1,\ldots,x_s),\quad s\geq1,
\end{eqnarray}
where on $\mathcal{D}\subset \mathcal{C}_s$ the operators $\mathcal{L}(j)$ and $\mathcal{L}_{\mathrm{int}}(j_1,j_{2})$
in a dimensionless form are defined by the formulas:
\begin{eqnarray}\label{com}
   &&\mathcal{L}(j)b_n\doteq \langle p_{j},\frac{\partial}{\partial q_{j}}\rangle b_n,\\
   &&\mathcal{L}_{\mathrm{int}}(j_1,j_{2})b_n\doteq \epsilon^{2}\int_{\mathbb{S}_+^2}d\eta\langle\eta,(p_{j_1}-p_{j_2})\rangle
     \big(b_n(x_1,\ldots,q_{j_1},p_{j_1}^*,\ldots,q_{j_2},p_{j_2}^*,\ldots,x_n)-\nonumber\\
   &&b_n(x_1,\ldots,x_n)\big)\delta(q_{j_1}-q_{j_2}+\epsilon\eta),\nonumber
\end{eqnarray}
respectively, and the coefficient $\epsilon>0$ is a scaling parameter (the ratio of the diameter
$\sigma>0$ to the mean free path of hard spheres). If $t\leq0$, a generator of the dual BBGKY hierarchy
is determined by the corresponding expression \cite{GG18}.

We refer to recurrence evolution equations (\ref{dh}) as the dual BBGKY hierarchy for hard spheres.
Let us adduce the simplest examples of recurrence evolution equations (\ref{dh}):
\begin{eqnarray*}
   &&\hskip-9mm\frac{\partial}{\partial t}B_{1}(t,x_1)=\mathcal{L}(1)B_{1}(t,x_1), \\
   &&\hskip-9mm\frac{\partial}{\partial t}B_{2}(t,x_1,x_2)=\big(\sum\limits_{j=1}^{2}\mathcal{L}(j)+
      \mathcal{L}_{\mathrm{int}}(1,2)\big)B_{2}(t,x_1,x_2)+
      \mathcal{L}_{\mathrm{int}}(1,2)\big(B_{1}(t,x_1)+B_{1}(t,x_2)\big),
\end{eqnarray*}
where the operators $\mathcal{L}(j)$ and $\mathcal{L}_{\mathrm{int}}(1,2)$ are defined by (\ref{com}).
For the microscopic phase space densities, the dual BBGKY hierarchy was considered in the paper \cite{GShZ}.

Let $Y\equiv(1,\ldots,s), Z\equiv (j_1,\ldots,j_{n})\subset Y$ and $\{Y\setminus Z\}$ is the set
consisting of one element $Y\setminus Z=(1,\ldots,j_{1}-1,j_{1}+1,\ldots,j_{n}-1,j_{n}+1,\ldots,s)$, and
we introduce the declusterization mapping $\theta$ defined by the formula: $\theta(\{Y\setminus Z\},Z)=Y$.

On the space $C_{\gamma}$ of sequences $b=(b_0,b_1,\ldots,b_n,\ldots)$ of functions $b_n\in C_n$
equipped with the norm: $\|b\|_{C_{\gamma}}=\max_{n\geq 0}\,\frac{\gamma^{n}}{n!}\,\|b_n\|$,
for the abstract Cauchy problem (\ref{dh}),(\ref{dhi}) the following statement is true.
The solution $B(t)=(B_{0},B_{1}(t,x_1),\ldots,B_{s}(t,x_1,\ldots,x_s),\ldots)$ of the Cauchy problem
(\ref{dh}),(\ref{dhi}) is determined by the following expansions:
\begin{eqnarray}\label{sdh}
   &&\hskip-7mm B_{s}(t,x_1,\ldots,x_s)=\sum_{n=0}^s\,\frac{1}{n!}\sum_{j_1\neq\ldots\neq j_{n}=1}^s
      \mathfrak{A}_{1+n}\big(t,\{Y\setminus Z\},Z\big)\,
      B_{s-n}^{\epsilon,0}(x_1,\ldots,\\
   &&x_{j_1-1},x_{j_1+1},\ldots,x_{j_n-1},x_{j_n+1},\ldots,x_s),\quad s\geq1,\nonumber
\end{eqnarray}
where the generating operator of expansion (\ref{sdh}) is $(1+n)th$-order cumulant (\ref{cumulan}) which
is defined by the formula
\begin{eqnarray}\label{cumulant}
    &&\hskip-5mm \mathfrak{A}_{1+n}(t,\{Y\setminus Z\},Z)\doteq\sum\limits_{\mathrm{P}:\,(\{Y\setminus Z\}, Z)={\bigcup}_i X_i}
       (-1)^{\mathrm{|P|}-1}({\mathrm{|P|}-1})!\prod_{X_i\subset \mathrm{P}}S_{|\theta(X_i)|}(t,\theta(X_i)),
\end{eqnarray}
and notations accepted in formula (\ref{cumulan}) are used.
For $B(0)=(B_{0},B_{1}^{\epsilon,0},\ldots,B_{s}^{\epsilon,0},\ldots)\in C_{\gamma}^0\subset C_{\gamma}$
of finite sequences of infinitely differentiable functions with compact supports it is a classical
solution and for arbitrary initial data $B(0)\in C_{\gamma}$ it is a generalized solution.

The simplest examples of marginal observables \eqref{sdh} are given by the following expansions:
\begin{eqnarray*}
   &&B_{1}(t,x_1)=\mathfrak{A}_{1}(t,1)B_{1}^{\epsilon,0}(x_1),\\
   &&B_{2}(t,x_1,x_2)=\mathfrak{A}_{1}(t,\{1,2\})B_{2}^{\epsilon,0}(x_1,x_2)+
      \mathfrak{A}_{2}(t,1,2)(B_{1}^{\epsilon,0}(x_1)+B_{1}^{\epsilon,0}(x_2)).
\end{eqnarray*}

We emphasize that expansion (\ref{sdh}) can be also represented in the form of the perturbation
(iteration) series \cite{BGer} as a result of applying of analogs of the Duhamel equation to cumulants
(\ref{cumulant}) of groups of operators (\ref{Sspher}).

We note that one component sequences of marginal observables correspond to observables of certain structure,
namely the marginal observable $B^{(1)}=(0,b_{1}^{\epsilon}(x_1),0,\ldots)$ corresponds to the additive-type
observable, and the marginal observable $B^{(k)}=(0,\ldots,0,b_{k}^{\epsilon}(x_1,\ldots,x_k),0,\ldots)$
corresponds to the $k$-ary-type observable \cite{BGer}. If in capacity of initial data (\ref{dhi}) we consider
the additive-type marginal observable, then the structure of solution expansion \eqref{sdh} is simplified and
attains the form
\begin{eqnarray}\label{af}
     &&B_{s}^{(1)}(t,x_1,\ldots,x_s)=\mathfrak{A}_{s}(t,1,\ldots,s)\sum_{j=1}^s b_{1}^{\epsilon}(x_j), \quad s\geq 1,
\end{eqnarray}
where $\mathfrak{A}_{s}(t)$ is the $sth$-order cumulant (\ref{cumulan}) of groups of operators (\ref{Sspher}).
In the case of initial $k$-ary-type marginal observables solution expansion (\ref{sdh}) takes the form
\begin{eqnarray}\label{af-k}
     &&\hskip-5mm B_{s}^{(k)}(t,x_1,\ldots,x_s)=\frac{1}{(s-k)!}\sum_{j_1\neq\ldots\neq j_{s-k}=1}^s
      \mathfrak{A}_{1+s-k}\big(t,\{(1,\ldots,s)\setminus (j_1,\ldots,j_{s-k})\},\\
     &&\hskip+12mm j_1,\ldots,j_{s-k}\big)\,
      b_{k}^{\epsilon}(x_1,\ldots,x_{j_1-1},x_{j_1+1},\ldots,x_{j_s-k-1},x_{j_s-k+1},\ldots,x_s),\quad s\geq k,\nonumber
\end{eqnarray}
and, if $1\leq s<k$, we have: $B_{s}^{(k)}(t)=0$.

\textcolor{blue!50!black}{\subsection{The BBGKY hierarchy for hard spheres}}
The state of a hard sphere system of a non-fixed, i.e. arbitrary but finite, number of identical
particles (nonequilibrium grand canonical ensemble) is described by a sequence of the marginal
distribution functions $F_s(x_1,\ldots,x_s),\, s\geq1,$ which are symmetric with respect to the
permutations of the arguments $x_1,\ldots,x_s$ and equal to zero on a set of forbidden configurations.

The evolution of all possible states is described by the sequence of marginal distribution functions
$F_s(t,x_1,\ldots,x_s),\,s\geq1,$ governed by the BBGKY hierarchy (the dual hierarchy to the recurrence
evolution equations (\ref{dh})) \cite{CGP97}:
\begin{eqnarray}
  \label{NelBog1}
    &&\frac{\partial}{\partial t}F_s(t)=\mathcal{L}^\ast_{s}F_{s}(t)+
       \sum_{i=1}^{s}\int_{\mathbb{R}^{3}\times\mathbb{R}^{3}}dx_{s+1}
       \mathcal{L}^\ast_{\mathrm{int}}(i,s+1)F_{s+1}(t),\\ \nonumber\\
  \label{NelBog2}
    &&F_s(t)_{\mid t=0}=F_s^{\epsilon,0}, \quad s\geq1,
\end{eqnarray}
where $\epsilon>0$ is a scaling parameter (the ratio of the diameter $\sigma>0$ to the mean free
path of hard spheres). If $t>0$, in the generator of the hierarchy of evolution equations (\ref{NelBog1})
the operator $\mathcal{L}^{\ast}_{s}$ is defined by the Poisson bracket of noninteracting particles with
the corresponding boundary conditions on $\partial\mathbb{W}_{s}$ \cite{CGP97}:
\begin{eqnarray*}\label{OperL}
   &&\mathcal{L}^{\ast}_{s}F_{s}(t)\doteq-\sum\limits_{i=1}^{s}
     \langle p_{i},\frac{\partial}{\partial q_{i}}\rangle_{\mid_{\partial \mathbb{W}_{s}}}F_{s}(t,x_1,\dots,x_s),
\end{eqnarray*}
where $\langle\eta,(p_i-p_{s+1})\rangle\doteq{\sum}_{\alpha=1}^3\eta^{\alpha}(p_i^\alpha-p_{s+1}^\alpha)$
is a scalar product. Hence this operator is the generator of the Liouville equation for states and it
is an adjoint operator to the generator $\mathcal{L}_{s}$ of the Liouville equation for observables \cite{CGP97}.
For $t>0$ the operator $\mathcal{L}^\ast_{\mathrm{int}}(i,s+1)$ in a dimensionless form is determined
by the expression
\begin{eqnarray*}\label{aLint}
   &&\hskip-12mm \sum_{i=1}^{s}\int_{\mathbb{R}^{3}\times\mathbb{R}^{3}}dx_{s+1}
     \mathcal{L}^{\ast}_{\mathrm{int}}(i,s+1)F_{s+1}(t)=
     \epsilon^{2}\sum\limits_{i=1}^s\int_{\mathbb{R}^3\times\mathbb{S}_{+}^{2}}
     d p_{s+1}d\eta\,\langle\eta,(p_i-p_{s+1})\rangle\times\\
   &&\hskip-12mm \big(F_{s+1}(t,x_1,\ldots,q_i,p_i^{*},\ldots,x_s,q_i-\epsilon\eta,p_{s+1}^{*})
     -F_{s+1}(t,x_1,\ldots,x_s,q_i+\epsilon\eta,p_{s+1})\big),\nonumber
\end{eqnarray*}
where ${\Bbb S}_{+}^{2}\doteq\{\eta\in\mathbb{R}^{3}\big|\,|\eta|=1,\langle\eta,(p_i-p_{s+1})\rangle>0\}$
and the momenta $p_{i}^{*},p_{s+1}^{*}$ are defined by equalities (\ref{momenta}). If $t\leq0$, a generator
of the BBGKY hierarchy is determined by the corresponding operators \cite{CGP97}.

Subsequently, the initial data (\ref{NelBog2}) satisfying the chaos condition will be examined
\cite{CGP97}, i.e. we consider the initial state of statistically independent hard spheres
on allowed configurations
\begin{eqnarray}\label{Bog2_haos}
   &&F^{\epsilon,0}_{s}(x_1,\ldots,x_{s})
     =\prod_{i=1}^{s}F_{1}^{\epsilon,0}(x_i)\mathcal{X}(\mathbb{R}^{3s}\setminus \mathbb{W}_s), \quad s\geq1,
\end{eqnarray}
where $\mathcal{X}(\mathbb{R}^{3s}\setminus \mathbb{W}_s)$ is the Heaviside step function of the
allowed configurations $\mathbb{R}^{3s}\setminus \mathbb{W}_s$.

If a one-particle distribution function $F_{1}^{\epsilon,0}$ is an integrable function, then a global in time
solution of the Cauchy problem (\ref{NelBog1}),(\ref{Bog2_haos}) is determined by a sequence of the following
series expansions \cite{GRS04}:
\begin{eqnarray}\label{F(t)}
    &&\hskip-12mm F_{s}(t,x_1,\ldots,x_{s})=\\
    &&\hskip-12mm\sum\limits_{n=0}^{\infty}\frac{1}{n!}
        \int_{(\mathbb{R}^{3}\times\mathbb{R}^{3})^{n}}dx_{s+1}\ldots dx_{s+n}
        \mathfrak{A}_{1+n}^\ast(t,\{Y\},X\setminus Y)\prod_{i=1}^{s+n}F_{1}^{\epsilon,0}(x_i)
        \mathcal{X}(\mathbb{R}^{3(s+n)}\setminus \mathbb{W}_{s+n}),\quad s\geq1,\nonumber
\end{eqnarray}
where the generating operator $\mathfrak{A}_{1+n}^{\ast}(t)$ of series expansion
(\ref{F(t)}) is the $(n+1)th$-order cumulant of the groups of adjoint operators (\ref{ad})
which is defined by the following expansion
\begin{eqnarray}\label{nLkymyl}
   &&\hskip-8mm \mathfrak{A}_{1+n}^{\ast}(t,\{Y\},X\setminus Y)=
      \sum\limits_{\texttt{P}:\,(\{Y\},X\setminus Y)={\bigcup\limits}_i X_i}
      (-1)^{|\texttt{P}|-1}(|\texttt{P}|-1)!\prod_{X_i\subset \texttt{P}}S_{|\theta(X_i)|}^{\ast}(t,\theta(X_i)),
\end{eqnarray}
and the following notations are used: $\{Y\}$ is a set consisting of one element
$Y\equiv(1,\ldots,s)$, i.e. $|\{Y\}|=1$, $\sum_\texttt{P}$ is a sum over all possible
partitions $\texttt{P}$ of the set $(\{Y\},X\setminus Y)\equiv(\{Y\},s+1,\ldots,s+n)$
into $|\texttt{P}|$ nonempty mutually disjoint subsets $X_i\in(\{Y\},X\setminus Y)$,
the mapping $\theta$ is the declusterization mapping defined by the formula: $\theta(\{Y\},X\setminus Y)=X$.

We give the simplest examples of cumulants (\ref{nLkymyl}) of the groups of adjoint operators (\ref{ad}):
\begin{eqnarray*}
    &&\mathfrak{A}_{1}^{\ast}(t,\{Y\})\doteq S_{s}^{\ast}(t,1,\ldots,s),\\
    &&\mathfrak{A}_{1+1}^{\ast}(t,\{Y\},s+1)\doteq S_{s+1}^{\ast}(t,1,\ldots,s+1) - S_{s}^{\ast}(t,1,\ldots,s)S_{1}^{\ast}(t,s+1),\\
    &&\mathfrak{A}_{1+2}^{\ast}(t,\{Y\},s+1,s+2)\doteq S_{s+3}^{\ast}(t,1,\ldots,s+3) - S_{s+1}^{\ast}(t,1,\ldots,s+1)S_{1}^{\ast}(t,s+3)-\\
    &&\hskip+12mm S_{s+1}^{\ast}(t,1,\ldots,s,s+3)S_{1}^{\ast}(t,s+2) - S_{s}^{\ast}(t,1,\ldots,s)S_{2}^{\ast}(t,s+2,s+3)+ \\
    &&\hskip+12mm 2!S_{s}^{\ast}(t,1,\ldots,s)S_{1}^{\ast}(t,s+2)S_{1}^{\ast}(t,s+3).
\end{eqnarray*}

If a one-particle distribution function $F_{1}^{\epsilon,0}$ is a continuously differentiable integrable
function with compact support, solution (\ref{F(t)}) of the Cauchy problem (\ref{NelBog1}),(\ref{Bog2_haos})
is a strong solution and for arbitrary integrable functions $F_{1}^{\epsilon,0}$ it is a weak solution \cite{CGP97}.

We emphasize that series expansion (\ref{F(t)}) can be also represented in the form of the perturbation
(iteration) series as a result of applying of analogs of the Duhamel equation to cumulants (\ref{nLkymyl})
of the groups of operators (\ref{ad}). In the case of a system of infinitely many hard spheres \cite{GP85}
a solution of the Cauchy problem of the BBGKY hierarchy is determined by perturbation series for initial
data from the space $L^{\infty}_\xi$ of sequences of bounded functions equipped with the norm:
$\|f\|_{L^{\infty}_\xi}=\sup_{n\geq0}\xi^{-n}\sup_{x_1,\ldots,x_n}|f_n(x_1,\ldots,x_n)|\exp\big(\beta\sum_{i=1}^n\frac{p_i^2}{2}\big)$.
In this case the existence of local in time solution was proved in \cite{GP85} (see also \cite{PG90}).

We remark that for $t>0$ in a one-dimensional space, i.e. for gas of hard rods, operator (\ref{bLint})
has the form \cite{PG83}:
\begin{eqnarray*}\label{aLint1}
    &&\hskip-8mm\sum_{i=1}^{s}\int_{\mathbb{R}^{3}\times\mathbb{R}^{3}}dx_{s+1}
       \mathcal{L}^*_{\mathrm{int}}(i,s+1)F_{s+1}(t)=\\
    &&\hskip-8mm\sum\limits_{i=1}^s\int_{0}^{\infty}d P P\big(F_{s+1}(t,x_1,\ldots,q_i,p_i-P,\ldots,x_s,q_i-\epsilon,p_i)-
       F_{s+1}(t,x_1,\ldots,x_s,q_i-\epsilon,p_i+P)+ \nonumber \\
    &&F_{s+1}(t,x_1,\ldots,q_i,p_i+P,\ldots,x_s,q_i+\epsilon,p_i)-F_{s+1}(t,x_1,\ldots,x_s,q_i+\epsilon,p_i-P)\big),\nonumber
\end{eqnarray*}
and for $t<0$ this operator has the corresponding form \cite{CGP97}. In this case for initial data from the space
$L^{\infty}_\xi$ the existence of global in time solution was proved in \cite{Ger}.

\bigskip
\textcolor{blue!50!black}{\section{The kinetic evolution of observables}}

\vskip-5mm
\textcolor{blue!50!black}{\subsection{The Boltzmann--Grad limit theorem for observables}}
Let us consider the problem of a rigorous description of the kinetic evolution of a hard sphere
system in the framework of the dynamics of observables on the basis of the Boltzmann--Grad asymptotic
behavior of the stated above solution (\ref{sdh}) of the dual BBGKY hierarchy (\ref{dh}).

We will assume the existence of the Boltzmann--Grad scaling limit for initial data $B_{s}^{\epsilon,0}\in C_s$
in the following sense
\begin{eqnarray}\label{asumdino}
    &&\mathrm{w^{\ast}-}\lim\limits_{\epsilon\rightarrow 0}\big(\epsilon^{-2s} B_{s}^{\epsilon,0}-b_{s}^0\big)=0.
\end{eqnarray}
Then for arbitrary finite time interval there exists the Boltzmann--Grad limit of the
solution (\ref{sdh}) in the sense of the $\ast$-weak convergence of the space $C_s$
\begin{eqnarray}\label{asymto}
    &&\mathrm{w^{\ast}-}\lim\limits_{\epsilon\rightarrow 0}\big(\epsilon^{-2s}B_{s}(t)-b_{s}(t)\big)=0,
\end{eqnarray}
and it is defined by the expansion
\begin{eqnarray}\label{Iterd}
   &&\hskip-8mm b_{s}(t,x_1,\ldots,x_s)=
      \sum\limits_{n=0}^{s-1}\,\int_0^tdt_{1}\ldots\int_0^{t_{n-1}}dt_{n}
      \, S_{s}^{0}(t-t_{1})\sum\limits_{i_{1}\neq j_{1}=1}^{s}
      \mathcal{L}_{\mathrm{int}}^0(i_{1},j_{1})\,S_{s-1}^{0}(t_{1}-t_{2})\ldots
\end{eqnarray}
\begin{eqnarray}
   &&\hskip-8mm S_{s-n+1}^{0}(t_{n-1}-t_{n})
      \sum\limits^{s}_{\mbox{\scriptsize $\begin{array}{c}i_{n}\neq j_{n}=1,\\
      i_{n},j_{n}\neq (j_{1},\ldots,j_{n-1})\end{array}$}}\mathcal{L}_{\mathrm{int}}^0(i_{n},j_{n})
      S_{s-n}^{0}(t_{n})b_{s-n}^0((x_1,\ldots,x_s)\setminus(x_{j_{1}},\ldots,x_{j_{n}})),\nonumber
\end{eqnarray}
 where for the group of operators of noninteracting particles the following notation is used
\begin{eqnarray*}
   &&\hskip-8mm S_{s-n+1}^{0}(t_{n-1}-t_{n})
     \equiv S_{s-n+1}^{0}(t_{n-1}-t_{n},Y \setminus (j_{1},\ldots,j_{n-1}))
     =\prod\limits_{j\in Y \setminus (j_{1},\ldots,j_{n-1})}S_{1}(t_{n-1}-t_{n},j),
\end{eqnarray*}
and we denote by $\mathcal{L}_{\mathrm{int}}^{0}(j_1,j_{2})$ the operator
\begin{eqnarray}\label{int0}
   &&\mathcal{L}_{\mathrm{int}}^{0}(j_1,j_{2})b_n\doteq \int_{\mathbb{S}_+^2}d\eta\langle\eta,(p_{j_1}-p_{j_2})\rangle
     \big(b_n(x_1,\ldots,q_{j_1},p_{j_1}^*,\ldots,q_{j_2},p_{j_2}^*,\ldots,x_n)-\nonumber\\
   &&\hskip+25mm b_n(x_1,\ldots,x_n)\big)\delta(q_{j_1}-q_{j_2}).\nonumber
\end{eqnarray}

If $b^0\in C_{\gamma}$, then the sequence $b(t)=(b_0,b_1(t),\ldots,b_{s}(t),\ldots)$
of limit marginal observables (\ref{Iterd}) is a generalized global solution of the
Cauchy problem of the dual Boltzmann hierarchy with hard spheres collisions \cite{GG18}:
\begin{eqnarray}\label{vdh}
   &&\frac{\partial}{\partial t}b_{s}(t,x_1,\ldots,x_s)=\sum\limits_{j=1}^{s}\mathcal{L}(j)\,b_{s}(t,x_1,\ldots,x_s)+\\
   &&\sum_{j_1\neq j_{2}=1}^s\mathcal{L}_{\mathrm{int}}^{0}(j_1,j_{2})\,b_{s-1}(t,(x_1,\ldots,x_s)\setminus(x_{j_1})),\nonumber\\
\nonumber\\
\label{vdhi}
   &&b_{s}(t,x_1,\ldots,x_s)\mid_{t=0}=b_{s}^0(x_1,\ldots,x_s),\quad s\geq1.
\end{eqnarray}
It should be noted that equations set (\ref{vdh}) has the structure of recurrence evolution equations.
We make a few examples of the dual Boltzmann hierarchy (\ref{vdh})
\begin{eqnarray*}
    &&\frac{\partial}{\partial t}b_{1}(t,x_1)=\langle p_1,\frac{\partial}{\partial q_{1}}\rangle\,b_{1}(t,x_1),\\
    &&\frac{\partial}{\partial t}b_{2}(t,x_1,x_2)=
      \sum\limits_{j=1}^{2}\langle p_j,\frac{\partial}{\partial q_{j}}\rangle\,b_{2}(t,x_1,x_2)+\\
    &&\hskip+12mm \int_{\mathbb{S}_+^2}d\eta\langle\eta,(p_{1}-p_{2})\rangle
     \big(b_1(q_{1},p_{1}^*)-b_1(x_1)+b_1(q_{2},p_{2}^*)-b_1(x_2)\big)\delta(q_{1}-q_{2}).
\end{eqnarray*}

Let us consider a particular case of observables, namely the Boltzmann--Grad limit of the
additive-type marginal observables $B^{(1)}(0)=(0,b_{1}^{\epsilon}(x_1),0,\ldots)$.
In this case solution (\ref{sdh}) of the dual BBGKY hierarchy (\ref{dh}) has form (\ref{af}).
If for the additive-type marginal observables $B^{(1)}(0)$ condition (\ref{asumdino}) is satisfied, i.e.
\begin{equation*}
    \mathrm{w^{\ast}-}\lim\limits_{\epsilon\rightarrow 0}\big( \epsilon^{-2} b_{1}^{\epsilon}-b_{1}^{0}\big)=0,
\end{equation*}
then, according to statement (\ref{asymto}), for  marginal observables (\ref{af}) it holds
\begin{equation*}
    \mathrm{w^{\ast}-}\lim\limits_{\epsilon\rightarrow 0} \big(\epsilon^{-2s} B_{s}^{(1)}(t)-b_{s}^{(1)}(t)\big)=0,
\end{equation*}
where the limit function $b_{s}^{(1)}(t)$ is determined by the expression
\begin{eqnarray}\label{itvad}
   &&\hskip-10mm b_{s}^{(1)}(t,x_1,\ldots,x_s)=\int_0^t dt_{1}\ldots\int_0^{t_{s-2}}dt_{s-1}\,
       S_{s}^{0}(t-t_{1})\sum\limits_{i_{1}\neq j_{1}=1}^{s}
       \mathcal{L}_{\mathrm{int}}^0(i_{1},j_{1})\times \\
   &&S_{s-1}^{0}(t_{1}-t_{2})\ldots S_{2}^{0}(t_{s-2}-t_{s-1})\hskip-5mm
       \sum\limits^{s}_{\mbox{\scriptsize $\begin{array}{c}i_{s-1}\neq j_{s-1}=1,\\
       i_{s-1},j_{s-1}\neq (j_{1},\ldots,j_{s-2})\end{array}$}}\hskip-5mm
       \mathcal{L}_{\mathrm{int}}^0(i_{s-1},j_{s-1})\times \nonumber\\
   &&S_{1}^{0}(t_{s-1})\,b_{1}^{0}((x_1,\ldots,x_s)
       \setminus (x_{j_{1}},\ldots,x_{j_{s-1}})),\quad s\geq1.\nonumber
\end{eqnarray}
as a special case of expansion (\ref{Iterd}). We will give examples of expressions (\ref{itvad}):
\begin{eqnarray*}
   &&b_{1}^{(1)}(t,x_1)=S_{1}(t,1)\,b_{1}^{0}(x_1),\\
   &&b_{2}^{(1)}(t,x_1,x_2)=\int_0^t dt_{1}\prod\limits_{i=1}^{2}S_{1}(t-t_{1},i)\,
      \mathcal{L}_{\mathrm{int}}^0(1,2)\sum\limits_{j=1}^{2}S_{1}(t_{1},j)\,b_{1}^{0}(x_j).
\end{eqnarray*}

If for the initial $k$-ary-type marginal observable $b_{k}^{\epsilon}$ condition (\ref{asumdino})
is satisfied, i.e.
\begin{eqnarray*}
    &&\hskip-5mm \mathrm{w^{\ast}-}\lim\limits_{\epsilon\rightarrow 0}\big(\epsilon^{-2}
        b_{k}^{\epsilon}-b_{k}^{0}\big)=0,
\end{eqnarray*}
then, according to statement (\ref{asymto}), for $k$-ary-type marginal observables (\ref{af-k})
we derive
\begin{eqnarray*}
    &&\hskip-5mm \mathrm{w^{\ast}-}\lim\limits_{\epsilon\rightarrow 0}
                 \big(\epsilon^{-2s}B_{s}^{(k)}(t)-b_{s}^{(k)}(t)\big)=0,
\end{eqnarray*}
where the limit marginal observable $b_{s}^{(k)}(t)$ is determined as a special case
of expansion (\ref{Iterd}):
\begin{eqnarray}\label{kIterd}
   &&\hskip-12mm b_{s}^{(k)}(t,x_1,\ldots,x_s)=\int_0^tdt_{1}\ldots\int_0^{t_{s-k-1}}dt_{s-k}
      \prod\limits_{j\in(1,\ldots,s)}S_{1}(t-t_{1},j)\sum\limits_{i_{1}\neq j_{1}=1}^{s}
      \mathcal{L}_{\mathrm{int}}^0(i_{1},j_{1})\times\\
   &&\hskip-7mm \prod\limits_{j\in(1,\ldots,s)\setminus (j_{1})}S_{1}(t_{1}-t_{2},j)
      \ldots\prod\limits_{j\in(1,\ldots,s)\setminus(j_{1},\ldots,j_{s-k-1})}S_{1}(t_{s-k-1}-t_{s-k},j)\times\nonumber \\
   &&\hskip-7mm \hskip-2mm\sum\limits^{s}_{\mbox{\scriptsize $\begin{array}{c}i_{s-k}\neq j_{s-k}=1,\\
      i_{s-k},j_{s-k}\neq (j_{1},\ldots,j_{s-k-1})\end{array}$}}\hskip-2mm\mathcal{L}_{\mathrm{int}}^0(i_{s-k},j_{s-k})
      \prod\limits_{j\in(1,\ldots,s)\setminus(j_{1},\ldots,j_{s-k})}S_{1}(t_{s-k},j)
      b_{k}^0((x_1,\ldots,\nonumber \\
   &&\hskip-7mm x_s)\setminus(x_{j_{1}},\ldots,x_{j_{s-k}})),\quad 1\leq s\leq k.\nonumber
\end{eqnarray}

Thus, in the Boltzmann--Grad scaling limit the collective behavior (the kinetic evolution) of hard
spheres is described in terms of limit marginal observables (\ref{itvad}),(\ref{kIterd}) governed
by the dual Boltzmann hierarchy (\ref{vdh}).

\textcolor{blue!50!black}{\subsection{The relationship of the kinetic evolution of observables and states}}
Let us examine the relationship the constructed Boltzmann--Grad asymptotic behavior of marginal
observables with the nonlinear Boltzmann kinetic equation with hard sphere collisions \cite{GG18}.

We shall consider initial states satisfying a chaos condition (\ref{Bog2_haos}), which means the
absence of correlations among hard spheres at the initial time, i.e.
\begin{eqnarray}\label{h2}
    &&F^{(c)}\equiv \big(1,F_1^{\epsilon,0}(x_1),\ldots,\prod_{i=1}^s F_1^{\epsilon,0}(x_i)
        \mathcal{X}_{\mathbb{R}^{3s}\setminus \mathbb{W}_s},\ldots\big).
\end{eqnarray}
Such an assumption about the initial state is intrinsic for the kinetic description of gas because
in this case all possible states are specified only by means of a one-particle distribution function.

Let $F_1^{\epsilon,0}\in L^{\infty}_\xi(\mathbb{R}^3\times\mathbb{R}^3)$, i.e. the inequality holds:
$|F_1^{\epsilon,0}(x_i)|\leq \xi\exp(-\beta\frac{p^2_i}{2})$, where $\xi>0,\beta\geq0$ are parameters.
We assume that the Boltzmann--Grad limit of the initial one-particle distribution function
$F_{1}^{\epsilon,0}\in L^{\infty}_\xi(\mathbb{R}^3\times\mathbb{R}^3)$ exists in the sense of a weak
convergence of the space $L^{\infty}_\xi(\mathbb{R}^3\times\mathbb{R}^3)$, namely,
\begin{eqnarray}\label{lh2}
 &&\mathrm{w-}\lim_{\epsilon\rightarrow 0}(\epsilon^2\,F_{1}^{\epsilon,0}-f_{1}^0)=0,
\end{eqnarray}
then the Boltzmann--Grad limit of sequence (\ref{h2}) satisfies a chaos condition too, i.e.
$f^{(c)}\equiv\big(1,f_1^0(x_1),\ldots,\prod_{i=1}^{s}f_{1}^0(x_i),\ldots\big)$.

We note that assumption (\ref{lh2}) with respect the Boltzmann--Grad limit of initial states
holds true for the equilibrium state \cite{GP88}.

If $b(t)\in\mathcal{C}_{\gamma}$ and $|f_1^{0}(x_i)|\leq \xi\exp(-\beta\frac{p^2_i}{2})$, then
the Boltzmann--Grad limit of mean value functional (\ref{avmar-1}) exists under the condition
that \cite{PG90}:
$t<t_{0}\equiv\big(\text{const}(\xi,\beta)\|f_1^0\|_{L^{\infty}_\xi(\mathbb{R}^3\times\mathbb{R}^3)}\big)^{-1}$,
and it is determined by the following series expansion:
\begin{eqnarray*}
   &&\big(b(t),f^{(c)}\big)=\sum\limits_{s=0}^{\infty}\,\frac{1}{s!}\,
       \int_{(\mathbb{R}^{3}\times\mathbb{R}^{3})^{s}}
      dx_{1}\ldots dx_{s}\,b_{s}(t,x_1,\ldots,x_s)\prod\limits_{i=1}^{s} f_1^0(x_i).
\end{eqnarray*}

For this mean value functional in the case of the additive-type limit marginal observables (\ref{itvad}),
the following representation is true
\begin{eqnarray*}\label{avmar-2}
  &&\hskip-7mm\big(b^{(1)}(t),f^{(c)}\big) =\sum\limits_{s=0}^{\infty}\,\frac{1}{s!}\,
       \int_{(\mathbb{R}^{3}\times\mathbb{R}^{3})^{s}}
      dx_{1}\ldots dx_{s}\,b_{s}^{(1)}(t,x_1,\ldots,x_s)\prod \limits_{i=1}^{s}f_{1}^0(x_i)=\\
  &&\hskip+5mm \int_{\mathbb{R}^{3}\times\mathbb{R}^{3}}dx_{1}\,b_{1}^{0}(x_1)f_{1}(t,x_1),\nonumber
\end{eqnarray*}
where the function $b_{s}^{(1)}(t)$ is given by expansion (\ref{itvad}) and the one-particle distribution
function $f_{1}(t,x_1)$ is represented by the series
\begin{eqnarray}\label{viter1}
   &&\hskip-9mm f_{1}(t,x_1)=
        \sum\limits_{n=0}^{\infty}\int_0^tdt_{1}\ldots\int_0^{t_{n-1}}dt_{n}\,
        \int_{(\mathbb{R}^{3}\times\mathbb{R}^{3})^{n}}dx_{2}\ldots dx_{n+1}\,
        S_{1}^{\ast}(t-t_{1},1)\mathcal{L}_{\mathrm{int}}^{0,\ast}(1,2)\times\\
   &&\hskip-7mm \prod\limits_{j_1=1}^{2}S_{1}^{\ast}(t_{1}-t_{2},j_1)\ldots
        \prod\limits_{i_{n}=1}^{n}S_{1}^{\ast}(t_{n-1}-t_{n},i_{n})
        \sum\limits_{k_{n}=1}^{n}\mathcal{L}_{\mathrm{int}}^{0,\ast}(k_{n},n+1)
        \prod\limits_{j_n=1}^{n+1}S_{1}^{\ast}(t_{n},j_n)\prod\limits_{i=1}^{n+1}f_1^0(x_i).\nonumber
\end{eqnarray}
In series (\ref{viter1}) it was introduced the operator
\begin{eqnarray*}\label{aLint}
   &&\hskip-12mm\int_{\mathbb{R}^{3}\times\mathbb{R}^{3}}dx_{n+1}
     \mathcal{L}_{\mathrm{int}}^{0,\ast}(i,n+1)f_{n+1}(x_1,\ldots,x_{n+1})\equiv\\
   &&\hskip-12mm \int_{\mathbb{R}^3\times\mathbb{S}_{+}^{2}}d p_{n+1}d\eta\,\langle\eta,(p_i-p_{n+1})
      \rangle\big(f_{n+1}(x_1,\ldots,q_i,p_i^{*},\ldots,x_s,q_i,p_{n+1}^{*})-
      f_{n+1}(x_1,\ldots,x_s,q_i,p_{n+1})\big).\nonumber
\end{eqnarray*}
and the group of operators $S_{1}^{\ast}(t)$ is a group of adjoint operators (\ref{ad}) to operators (\ref{Sspher}).

The one-particle distribution function $f_{1}(t)$ is a solution of the Cauchy problem of the Boltzmann
kinetic equation with hard sphere collisions:
\begin{eqnarray}
  \label{Bolz}
    &&\hskip-8mm\frac{\partial}{\partial t}f_{1}(t,x_1)=
       -\langle p_1,\frac{\partial}{\partial q_1}\rangle f_{1}(t,x_1)+\\
    &&\int_{\mathbb{R}^3\times\mathbb{S}^2_+}d p_2\, d\eta
       \,\langle\eta,(p_1-p_2)\rangle\big(f_1(t,q_1,p_1^{*})f_1(t,q_1,p_2^{*})-
       f_1(t,x_1)f_1(t,q_1,p_2)\big), \nonumber\\ \nonumber\\
  \label{Bolzi}
    &&\hskip-8mm f_1(t,x_1)_{\mid t=0}=f_{1}^0(x_1),
\end{eqnarray}
where the momenta $p_{1}^*$ and $p_{2}^*$ are pre-collision momenta of hard spheres (\ref{momenta}).

Thus, the hierarchy of evolution equations (\ref{vdh}) for additive-type marginal observables and initial state
(\ref{lh2}) describe the evolution of a hard sphere system just as the Boltzmann kinetic equation (\ref{Bolz}).

Correspondingly, if the initial state of hard spheres specified by a sequence of marginal distribution
functions (\ref{h2}), then for the state in the Boltzmann--Grad limit the property of propagation of initial
chaos holds. It is a result of the validity of the following equality for representations of the mean value
functional in the case of $k$-ary limit marginal observables (\ref{kIterd})
\begin{eqnarray}\label{pchaos}
    &&\hskip-7mm\big(b^{(k)}(t),f^{(c)}\big)=\sum\limits_{s=0}^{\infty}\,\frac{1}{s!}\,
       \int_{(\mathbb{R}^{3}\times\mathbb{R}^{3})^{s}}
      dx_{1}\ldots dx_{s}\,b_{s}^{(k)}(t,x_1,\ldots,x_s) \prod \limits_{i=1}^{s} f_1^0(x_i)=\\
    &&\hskip+5mm\frac{1}{k!}\int_{(\mathbb{R}^{3}\times\mathbb{R}^{3})^{k}}
      dx_{1}\ldots dx_{k}\,b_{k}^{0}(x_1,\ldots,x_k)
       \prod\limits_{i=1}^{k}f_{1}(t,x_i),\quad k\geq2,\nonumber
\end{eqnarray}
where for finite time interval the limit one-particle distribution function $f_{1}(t)$ is
represented by series expansion (\ref{viter1}) and therefore it is a solution of the Cauchy problem
of the Boltzmann kinetic equation (\ref{Bolz}),(\ref{Bolzi}) with hard sphere collisions.

Thus, in the Boltzmann--Grad scaling limit an equivalent approach to the description of the kinetic
evolution of hard spheres by means of the Cauchy problem of the Boltzmann kinetic equation
(\ref{Bolz}),(\ref{Bolzi}) is given by the Cauchy problem for the dual Boltzmann hierarchy with hard
sphere collisions (\ref{vdh}),(\ref{vdhi}) for the additive-type marginal observables. In the case of
the nonadditive-type marginal observables, a solution of the dual Boltzmann hierarchy with hard sphere
collisions (\ref{vdh}) is equivalent to the property of the propagation of initial chaos for the state
in the sense of equality (\ref{pchaos}).

\textcolor{blue!50!black}{\subsection{Remarks on the Boltzmann--Grad limit theorem for states}}
Let us remind that for functional (\ref{avmar-1}) of the mean value of observables the following
equality holds:
\begin{eqnarray*}
     &&\big(B(t),F(0)\big)=\big(B(0),F(t)\big),
\end{eqnarray*}
where the sequence $B(t)=(B_0,B_{1}(t),\ldots,B_{n}(t),\ldots)$ is solution (\ref{sdh}) of the
dual BBGKY hierarchy for hard spheres (\ref{dh}) and the sequence $F(t)=(1,F_{1}(t),\ldots,F_{n}(t),\ldots)$
is solution (\ref{F(t)}) of the BBGKY hierarchy for hard spheres (\ref{NelBog1}). This equality
proves the equivalence of the description of evolution within the framework of observables and
within the framework of states.

Usually, a solution of the BBGKY hierarchy is constructed by methods of perturbation theory
\cite{CGP97},\cite{Sp91},\cite{CIP},\cite{L},\cite{B46}. A nonperturbative solution was constructed
in the space of sequences of integrable functions in the paper \cite{GRS04} (see also \cite{CGP97}).

In the case of a system of infinitely many hard spheres \cite{SR12}-\cite{PG90} a local in time solution
\cite{GP85} of the Cauchy problem of the BBGKY hierarchy is determined by perturbation series \cite{CGP97},
\cite{CIP},\cite{PG90},\cite{Sp06} for arbitrary initial data from the space $L^{\infty}_\xi$ of sequences
of bounded functions equipped with the norm:
$\|f\|_{L^{\infty}_\xi}=\sup_{n\geq0}\xi^{-n}\sup_{x_1,\ldots,x_n}|f_n(x_1,\ldots,x_n)|\exp\big(\beta\sum_{i=1}^n\frac{p_i^2}{2}\big)$.
In this case the existence of the mean value functionals $\big(B(0),F(t)\big)$ and $\big(B(t),F(0)\big)$
were proved in papers \cite{PG90} and \cite{R10}, respectively.

Recently, there has been unflagging interest in the problem of deriving kinetic equations from the dynamics
of many particles as an asymptotic behavior of the BBGKY hierarchy in the scaling limits \cite{APST}-\cite{D17}.
In particular, the progress in the rigorous solution this problem on the basis of perturbation theory for
a system of hard spheres was achieved in the Boltzmann--Grad limit in the works \cite{SR12}-\cite{PG90}.
The Boltzmann--Grad limit theorem for equilibrium states was proved in the paper \cite{GP88}.

We also note that the property of the decay of correlations \cite{B46} of a solution of the BBGKY hierarchy
for hard spheres (\ref{NelBog1}) was proved in \cite{GerSh}.

\bigskip
\textcolor{blue!50!black}{\section{Asymptotic behavior of a hard sphere system with initial correlations}}

\vskip-5mm
\textcolor{blue!50!black}{\subsection{The Boltzmann kinetic equation with initial correlations}}
Let us consider initial states of a hard sphere system specified by the one-particle distribution
function $F_1^{0,\epsilon}\in L^{\infty}_\xi(\mathbb{R}^3\times\mathbb{R}^3)$ in the presence of
correlations, namely the initial states specified by the following sequence of marginal distribution
functions
\begin{eqnarray}\label{ins}
   &&\hskip-8mm F^{(cc)}=\big(1,F_1^{0,\epsilon}(x_1),g_{2}^{\epsilon}\prod_{i=1}^{2}F_1^{0,\epsilon}(x_i),\ldots,
        g_{n}^{\epsilon}\prod_{i=1}^{n}F_1^{0,\epsilon}(x_i),\ldots\big),
\end{eqnarray}
where the functions
$g_{n}^{\epsilon}\equiv g_{n}^{\epsilon}(x_1,\ldots,x_n)\in C_n(\mathbb{R}^{3n}\times(\mathbb{R}^{3n}\setminus\mathbb{W}_n)),\,n\geq2$,
specify the initial correlations of a hard sphere system. Since many-particle systems in condensed states
are characterized by correlations, sequence (\ref{ins}) describes the initial state typical for the kinetic
evolution of hard sphere fluids \cite{GerUJP}.

We assume that the Boltzmann--Grad limit of the initial one-particle distribution function
$F_{1}^{0,\epsilon}\in L^{\infty}_\xi(\mathbb{R}^3\times\mathbb{R}^3)$ exists in the sense as
above, i.e. in the sense of weak convergence, the equality holds:
$\mathrm{w-}\lim_{\epsilon\rightarrow 0}(\epsilon^2\,F_{1}^{0,\epsilon}-f_{1}^0)=0$,
and in the case of correlation functions let be:
$\mathrm{w-}\lim_{\epsilon\rightarrow 0}(g_{n}^{\epsilon}-g_{n})=0,\,n\geq2$.
Then in the Boltzmann--Grad limit the initial state (\ref{ins}) is determined by the following
sequence of the limit marginal distribution functions
\begin{eqnarray}\label{lins}
   &&\hskip-8mm f^{(cc)}=\big(1,f_1^{0}(x_1),g_{2}\prod_{i=1}^{2}f_1^{0}(x_i),\ldots,
        g_{n}\prod_{i=1}^{n}f_1^{0}(x_i),\ldots\big).
\end{eqnarray}

Let us consider relationships of the constructed Boltzmann--Grad asymptotic behavior of marginal
observables with the nonlinear Boltzmann-type kinetic equation for initial states (\ref{lins}).

For the limit additive-type marginal observables (\ref{itvad}) and initial states (\ref{lins})
the following equality is true:
\begin{eqnarray*}\label{avmar-2}
  &&\hskip-7mm \big(b^{(1)}(t),f^{(cc)}\big)=\sum\limits_{s=0}^{\infty}\,\frac{1}{s!}\,
      \int_{(\mathbb{R}^{3}\times\mathbb{R}^{3})^{s}}dx_{1}\ldots dx_{s}
     \,b_{s}^{(1)}(t,x_{1},\ldots,x_{s})g_{s}(x_{1},\ldots,x_{s})\prod_{i=1}^{s}f_1^{0}(x_i)=\\
  &&\int_{\mathbb{R}^{3}\times\mathbb{R}^{3}}dx_{1}\,b_{1}^{0}(x_1)f_{1}(t,x_1),\nonumber
\end{eqnarray*}
where the functions $b_{s}^{(1)}(t),\,s\geq1,$ are represented by expansions (\ref{itvad}) and the
limit one-particle distribution function $f_{1}(t)$ is represented by the following series expansion
\begin{eqnarray}\label{viterc}
   &&\hskip-7mm f_{1}(t,x_1)=\\
   &&\hskip-5mm \sum\limits_{n=0}^{\infty}\,\int_0^tdt_{1}\ldots\int_0^{t_{n-1}}dt_{n}\,
        \int_{(\mathbb{R}^{3}\times\mathbb{R}^{3})^{n}}dx_{2}\ldots dx_{n+1}
        S_{1}^{\ast}(t-t_{1},1)\mathcal{L}_{\mathrm{int}}^{0,\ast}(1,2)S_{1}^{\ast}(t_{1}-t_{2},j_1)\ldots\nonumber\\
   &&\hskip-5mm \prod\limits_{i_{n}=1}^{n}S_{1}^{\ast}(t_{n}-t_{n},i_{n})
        \sum\limits_{k_{n}=1}^{n}\mathcal{L}_{\mathrm{int}}^{0,\ast}(k_{n},n+1)
        \prod\limits_{j_n=1}^{n+1}S_{1}^{\ast}(t_{n},j_n)g_{1+n}(x_1,\ldots,x_{n+1})\prod\limits_{i=1}^{n+1}f_1^0(x_i).\nonumber
\end{eqnarray}
Series (\ref{viterc}) is uniformly convergent for finite time interval under the condition as above (\ref{viter1}).

The function $f_{1}(t)$ represented by series (\ref{viterc}) is a weak solution of the Cauchy problem of
the Boltzmann kinetic equation with initial correlations \cite{G14},\cite{GK}
\begin{eqnarray}\label{Bc}
   &&\hskip-5mm\frac{\partial}{\partial t}f_{1}(t,x_1)=-\langle p_1,\frac{\partial}{\partial q_1}\rangle f_{1}(t,x_1)+\\
   &&+\int_{\mathbb{R}^3\times\mathbb{S}^2_+}d p_2\,d\eta
       \,\langle\eta,(p_1-p_2)\rangle\Big(g_{2}(q_1-p_1^{*}t,p_1^{*},q_2-p_2^{*}t,p_2^{*})f_1(t,q_1,p_1^{*})f_1(t,q_1,p_2^{*})
       -\nonumber \\
   &&-g_{2}(q_1-p_1t,p_1,q_2-p_2t,p_2)f_1(t,x_1)f_1(t,q_1,p_2)\Big), \nonumber\\ \nonumber\\
\label{Bci}
   &&\hskip-5mmf_{1}(t,x_1)|_{t=0}=f_{1}^0(x_1).
\end{eqnarray}

We emphasize that the Boltzmann equation with initial correlations (\ref{Bc}) is the non-Markovian kinetic
equation which describe memory effects in fluids of hard spheres.

Thus, in the case of initial states specified by one-particle distribution function (\ref{lins})
we establish that the dual Boltzmann hierarchy with hard sphere collisions (\ref{vdh}) for additive-type
marginal observables describes the evolution of a hard sphere system just as the Boltzmann kinetic equation
with initial correlations (\ref{Bc}).

\textcolor{blue!50!black}{\subsection{On the propagation of initial correlations}}
The property of the propagation of initial correlations in a low-density limit is a consequence of the validity
of the following equality for a mean value functional of the limit $k$-ary marginal observables:
\begin{eqnarray}\label{dchaos}
    &&\hskip-12mm \big(b^{(k)}(t),f^{(cc)}\big)=\sum\limits_{s=0}^{\infty}\,\frac{1}{s!}\,
       \int_{(\mathbb{R}^{3}\times\mathbb{R}^{3})^{s}}dx_{1}\ldots dx_{s}\,b_{s}^{(k)}(t,x_1,\ldots,x_s)
       g_{s}(x_1,\ldots,x_s)\prod \limits_{j=1}^{s} f_1^0(x_j)=\\
    &&\hskip-12mm \frac{1}{k!}\int_{(\mathbb{R}^{3}\times\mathbb{R}^{3})^{k}}dx_{1}\ldots dx_{k}
       \,b_{k}^{0}(x_1,\ldots,x_k)\prod_{i_1=1}^{k}S_{1}^{\ast}(t,i_1)g_{k}(x_1,\ldots,x_k)
       \prod_{i_2=1}^{k}(S_{1}^{\ast})^{-1}(t,i_2)\prod\limits_{j=1}^{k}f_{1}(t,x_j),\nonumber
\end{eqnarray}
where the one-particle marginal distribution function $f_{1}(t,x_j)$ is solution (\ref{viterc}) of the
Cauchy problem of the Boltzmann kinetic equation with initial correlations (\ref{Bc}),(\ref{Bci}), and
the inverse group to the group of operators
$S_{1}^{\ast}(t)$ we denote by $(S_{1}^{\ast})^{-1}(t)=S_{1}^{\ast}(-t)=S_{1}(t)$.

We note that, according to equality (\ref{dchaos}), in the Boltzmann--Grad limit the marginal
correlation functions defined as cluster expansions of marginal distribution functions, namely,
\begin{eqnarray*}
   &&f_{s}(t,x_1,\ldots,x_s)=
      \sum\limits_{\mbox{\scriptsize $\begin{array}{c}\mathrm{P}:(x_1,\ldots,x_s)=\bigcup_{i}X_{i}\end{array}$}}
      \prod\limits_{X_i\subset\mathrm{P}}g_{|X_i|}(t,X_i),\quad s\geq1,
\end{eqnarray*}
has the following explicit form:
\begin{eqnarray*}\label{cfmf}
    &&\hskip-5mm g_{1}(t,x_1)=f_{1}(t,x_1),\\
    &&\hskip-5mm g_{s}(t,x_1,\ldots,x_s)=
        \tilde{g}_{s}(q_1-p_1t,p_1,\ldots,q_s-p_st,p_s)\prod\limits_{j=1}^{s}f_{1}(t,x_j),\quad s\geq2,\nonumber
\end{eqnarray*}
where the one-particle distribution function $f_{1}(t)$ is a solution of the Cauchy problem of the Boltzmann
kinetic equation with initial correlations (\ref{Bc}),(\ref{Bci}) and for initial correlation functions of state
(\ref{lins}) it is used the following notation:
\begin{eqnarray*}\label{FG1}
   &&\hskip-5mm \tilde{g}_{s}(x_1,\ldots,x_s)=
      \sum\limits_{\mbox{\scriptsize$\begin{array}{c}\mathrm{P}:(x_1,\ldots,x_s)=\bigcup_{i}X_{i}\end{array}$}}
      \prod_{X_i\subset \mathrm{P}}g_{|X_i|}(X_i).
\end{eqnarray*}
In this expansion the symbol $\sum_\mathrm{P}$ means the sum over possible partitions $\mathrm{P}$ of the set of arguments
$(x_1,\ldots,x_s)$ on $|\mathrm{P}|$ nonempty subsets $X_i$ and the functions $g_{|X_i|},\,X_i\subset \mathrm{P}$ describe
limit correlations of state (\ref{lins}).

Thus, in the case of limit $k$-ary marginal observables a solution of the dual Boltzmann hierarchy
with hard sphere collisions (\ref{vdh}) is equivalent to a property of propagation of the initial
correlations for the $k$-particle distribution function in the sense of equality (\ref{dchaos})
or in other words the Boltzmann--Grad dynamics does not create new correlations except initial correlations.

\bigskip
\textcolor{blue!50!black}{\section{The generalized Enskog equation}}

\vskip-5mm
\textcolor{blue!50!black}{\subsection{On the origin of the kinetic evolution of states}}
In the case of initial state (\ref{Bog2_haos}) the dual picture of the evolution to the picture
of the evolution described employing observables of a system of hard spheres governed by the dual
BBGKY hierarchy (\ref{dh}) for the marginal observables is the evolution of states described by
means of the non-Markovian Enskog kinetic equation for the one-particle distribution function and
a sequence of explicitly defined functionals of the solution of such a kinetic equation that are
described the evolution of all possible correlations in a system of hard spheres \cite{GG11}.

In view of the fact that the initial state is completely specified by a one-particle
marginal distribution function on allowed configurations (\ref{Bog2_haos}), for mean value
functional (\ref{avmar-1}) the following representation holds \cite{GG}:
\begin{eqnarray}\label{w}
    &&\big(B(t), F^{(c)}\big)=\big(B(0), F(t\mid F_{1}(t))\big),
\end{eqnarray}
where $F^{(c)}$ is the sequence of initial marginal distribution functions (\ref{h2}), and the sequence
$F(t\mid F_{1}(t))=\big(1,F_1(t),F_2(t\mid F_{1}(t)),\ldots,F_s(t\mid F_{1}(t))\big)$
is a sequence of the marginal functionals of state $F_{s}(t,x_1,\ldots,x_s \mid F_{1}(t)),\, s\geq2,$
represented by the series expansions over the products with respect to the one-particle
distribution function $F_{1}(t)$, namely
\begin{eqnarray}\label{f}
   &&\hskip-12mm F_{s}(t,x_1,\ldots,x_s\mid F_{1}(t))\doteq\\
   &&\hskip-12mm \sum _{n=0}^{\infty }\frac{1}{n!}\,\int_{(\mathbb{R}^{3}\times\mathbb{R}^{3})^{n}}
      dx_{s+1}\ldots dx_{s+n}\,\mathfrak{V}_{1+n}(t,\{Y\},X\setminus Y)
      \prod_{i=1}^{s+n}F_{1}(t,x_i),\quad s\geq 2,\nonumber
\end{eqnarray}
where the following notations are used: $Y\equiv(1,\ldots,s), X\equiv(1,\ldots,s+n)$.
The generating operator of series (\ref{f}) is the $(1+n)th$-order operator $\mathfrak{V}_{1+n}(t),\,n\geq0$,
defined by the following expansion \cite{GG}:
\begin{eqnarray}\label{skrrn}
    &&\hskip-8mm\mathfrak{V}_{1+n}(t,\{Y\},X\setminus Y)\doteq\sum_{k=0}^{n}(-1)^k\,\sum_{m_1=1}^{n}\ldots
       \sum_{m_k=1}^{n-m_1-\ldots-m_{k-1}}\frac{n!}{(n-m_1-\ldots-m_k)!}\times\\
    &&\hskip-8mm \widehat{\mathfrak{A}}_{1+n-m_1-\ldots-m_k}(t,\{Y\},s+1,
       \ldots,s+n-m_1-\ldots-m_k)\prod_{j=1}^k\,\sum_{k_2^j=0}^{m_j}\ldots\nonumber\\
    &&\hskip-8mm\sum_{k^j_{n-m_1-\ldots-m_j+s}=0}^{k^j_{n-m_1-\ldots-m_j+s-1}}\,\prod_{i_j=1}^{s+n-m_1-\ldots-m_j}
       \frac{1}{(k^j_{n-m_1-\ldots-m_j+s+1-i_j}-k^j_{n-m_1-\ldots-m_j+s+2-i_j})!}\times\nonumber\\
    &&\hskip-8mm \widehat{\mathfrak{A}}_{1+k^j_{n-m_1-\ldots-m_j+s+1-i_j}-k^j_{n-m_1-\ldots-m_j+s+2-i_j}}(t,
       i_{j},s+n-m_1-\ldots-m_j+1 +\nonumber \\
    &&\hskip-8mm k^j_{s+n-m_1-\ldots-m_j+2-i_j},\ldots,s+n-m_1-\ldots-m_j+k^j_{s+n-m_1-\ldots-m_j+1-i_j}),\nonumber
\end{eqnarray}
where it means that: $k^j_1\equiv m_j,\,k^j_{n-m_1-\ldots-m_j+s+1}\equiv 0$, and
the $(1+n)th$-order scattering cumulant we denoted by the operator $\widehat{\mathfrak{A}}_{1+n}(t)$:
\begin{eqnarray*}\label{scacu}
   &&\hskip-8mm\widehat{\mathfrak{A}}_{1+n}(t,\{Y\},X\setminus Y)
      \doteq\mathfrak{A}_{1+n}^\ast(t,\{Y\},X\setminus Y)\mathcal{X}_{\mathbb{R}^{3(s+n)}\setminus\mathbb{W}_{s+n}}
      \prod_{i=1}^{s+n}\mathfrak{A}_{1}^\ast(t,i)^{-1},
\end{eqnarray*}
and the operator $\mathfrak{A}_{1+n}^\ast(t)$ is the $(1+n)th$-order cumulant of adjoint groups of operators
of hard spheres (\ref{nLkymyl}). We adduce some examples of expressions (\ref{skrrn}):
\begin{eqnarray*}
   &&\mathfrak{V}_{1}(t,\{Y\})=\widehat{\mathfrak{A}}_{1}(t,\{Y\})\doteq
       S_s^\ast(t,1,\ldots,s)\mathcal{X}_{\mathbb{R}^{3(s)}\setminus \mathbb{W}_{s}}\prod_{i=1}^{s}S_1^\ast(t,i)^{-1},\\
   &&\mathfrak{V}_{2}(t,\{Y\},s+1)=\widehat{\mathfrak{A}}_{2}(t,\{Y\},s+1)-
       \widehat{\mathfrak{A}}_{1}(t,\{Y\})\sum_{i_1=1}^s
       \widehat{\mathfrak{A}}_{2}(t,i_1,s+1).\nonumber
\end{eqnarray*}

We emphasize that marginal functionals of state (\ref{f}) describe the correlations generated by dynamics
of a hard sphere system.

The one-particle distribution function $F_{1}(t)$, i.e. the first element of the sequence $F(t\mid F_{1}(t))$,
is determined by series (\ref{F(t)}), namely
\begin{eqnarray}\label{F(t)1}
    &&\hskip-12mm F_{1}(t,x_1)=\sum\limits_{n=0}^{\infty}\frac{1}{n!}
      \int_{(\mathbb{R}^3\times\mathbb{R}^3)^n}dx_2\ldots dx_{n+1}\,
      \mathfrak{A}_{1+n}^\ast(t,1,\ldots,n+1)
      \mathcal{X}_{\mathbb{R}^{3(1+n)}\setminus\mathbb{W}_{1+n}}\prod_{i=1}^{n+1}F_{1}^{\epsilon,0}(x_i),
\end{eqnarray}
where the generating operator $\mathfrak{A}_{1+n}^\ast(t)$ is the $(1+n)$th-order cumulant (\ref{nLkymyl})
of the groups of adjoint operators (\ref{ad}).

A method of the construction of marginal functionals of state (\ref{f}) is based on the application
of the kinetic cluster expansions \cite{GerUJP} to the generating operators of series (\ref{F(t)})
(expressions (\ref{skrrn}) are solutions of recurrence relations representing such kinetic cluster
expansions).

Let us note that in particular case of initial data (\ref{dhi}) specified by the additive-type marginal
observables, according to solution expansion (\ref{af}), equality
(\ref{w}) takes the form
\begin{eqnarray}\label{avmar-11}
   &&\big( B^{(1)}(t),F(0)\big)=
      \int_{\mathbb{R}^{3}\times\mathbb{R}^{3}}dx_{1}\,b_{1}^{\epsilon}(x_1)F_{1}(t,x_1),\nonumber
\end{eqnarray}
where the one-particle distribution function $F_{1}(t)$ is determined by series (\ref{F(t)1}).
In the case of initial data (\ref{dhi}) specified by the $s$-ary marginal observable $s\geq2$,
equality (\ref{w}) has the form
\begin{eqnarray}\label{avmar-12}
   &&\hskip-5mm\big( B^{(s)}(t),F(0)\big)=\frac{1}{s!}\int_{(\mathbb{R}^{3}\times\mathbb{R}^{3})^{s}}
      dx_{1}\ldots dx_{s}\,b_{s}^{\epsilon}(x_1,\ldots,x_s)F_s(t,x_1,\ldots,x_s\mid F_{1}(t)),\nonumber
\end{eqnarray}
where the marginal functionals of state $F_{s}(t,x_1,\ldots,x_s \mid F_{1}(t))$ are determined
by series (\ref{f}).

Thus, for the initial state specified by the one-particle distribution function, the evolution
of all possible states of a system of many hard spheres can be described by means of the state
of a typical particle without any scaling approximations.

We remark also that in the case of the initial state that involves correlations (\ref{ins}) considered
approach permits to take into consideration the initial correlations in the kinetic equations \cite{G14}.

\textcolor{blue!50!black}{\subsection{The non-Markovian Enskog kinetic equation}}
For $t\geq 0$ the one-particle distribution function (\ref{F(t)1}) is a solution of the following
Cauchy problem of the non-Markovian generalized Enskog kinetic equation \cite{GG},\cite{GG11}:
\begin{eqnarray}
 \label{gke1}
   &&\hskip-12mm\frac{\partial}{\partial t}F_{1}(t,q_1,p_1)=
      -\langle p_1,\frac{\partial}{\partial q_1}\rangle F_{1}(t,q_1,p_1)+
\end{eqnarray}
\begin{eqnarray}
   &&\hskip-12mm\epsilon^2\int_{\mathbb{R}^3\times\mathbb{S}^2_+}d p_2 d\eta
      \langle\eta,(p_1-p_2)\rangle\Big(F_2(t,q_1,p_1^{*},q_1-\epsilon\eta,p_2^{*}\mid F_{1}(t))-\nonumber\\
   &&\hskip-12mm F_2(t,q_1,p_1,q_1+\epsilon\eta,p_2\mid F_{1}(t))\Big),\nonumber\\ \nonumber\\
 \label{gkei}
   &&\hskip-12mm F_{1}(t)|_{t=0}= F_{1}^{\epsilon,0},
\end{eqnarray}
where the collision integral is determined by the marginal functional of the state (\ref{f}) in the case
of $s=2$ and the expressions $p_1^{*}$ and $p_2^{*}$ are the pre-collision momenta of hard spheres (\ref{momenta}).

Hence in the case of the additive-type marginal observables the generalized Enskog kinetic equation (\ref{gke1})
is dual to the dual BBGKY hierarchy of hard spheres (\ref{dh}) with respect to bilinear form (\ref{avmar-1}).

We note that the structure of collision integral of the generalized Enskog equation (\ref{gke1}) is such that
the first term of its expansion is the collision integral of the Boltzman--Enskog kinetic equation and the next
terms describe the all possible correlations which are created by the dynamics of hard spheres and by the
propagation of initial correlations connected with the forbidden configurations, indeed
\begin{eqnarray*}\label{gke1e}
   &&\hskip-7mm\frac{\partial}{\partial t}F_{1}(t,x_1)=
                -\langle p_1,\frac{\partial}{\partial q_1}\rangle F_{1}(t,x_1)+\mathcal{I}_{GEE},
\end{eqnarray*}
where the collision integral is determined by the following series expansion:
\begin{eqnarray*}
   &&\hskip-7mm\mathcal{I}_{GEE}\doteq\epsilon^2\sum_{n=0}^{\infty}\frac{1}{n!}\int_{\mathbb{R}^3\times\mathbb{S}^2_+}d p_2 d\eta
      \int_{(\mathbb{R}^{3}\times\mathbb{R}^{3})^{n}}dx_3\ldots dx_{n+2}\langle\eta,(p_1-p_2)\rangle \times\nonumber\\
   &&\hskip-7mm\big(\mathfrak{V}_{1+n}(t,\{1^{*},2^{*}_{-}\},3,\ldots,n+2)
      F_1(t,q_1,p_1^{*})F_1(t,q_1-\epsilon\eta,p_2^{*})\prod_{i=3}^{n+2}F_{1}(t,x_i)- \nonumber\\
   &&\hskip-7mm\mathfrak{V}_{1+n}(t,\{1,2_{+}\},3,\ldots,n+2)F_1(t,x_1)
      F_1(t,q_1+\epsilon\eta,p_2)\prod_{i=3}^{n+2}F_{1}(t,x_i)\big),\nonumber
\end{eqnarray*}
and it was used the notations adopted to the conventional notation of
the Enskog collision integral: indices $(1^{\sharp},2^{\sharp}_{\pm})$ denote that the evolution operator
$\mathfrak{V}_{1+n}(t)$ acts on the corresponding phase points $(q_1,p_1^{\sharp})$ and $(q_1\pm\epsilon\eta,p_2^{\sharp})$,
and the $(n+1)$th-order evolution operator $\mathfrak{V}_{1+n}(t),\,n\geq0$, is determined by expansion
(\ref{skrrn}) in the case of $|Y|=2$. The series on the right-hand side of this equation converges under
the condition: $\|F_1(t)\|_{L^1(\mathbb{R}\times\mathbb{R})}<e^{-8}$ .

We remark that in the paper \cite{GG} for the initial-value problem (\ref{gke1}),(\ref{gkei}) the existence
theorem was proved in the space of integrable functions and the correspondence of the generalized Enskog
equation (\ref{gke1}) with the Markovian Enskog-type kinetic equations was established. In the paper \cite{T}
for kinetic equation (\ref{gke1}) it was established the explicit soliton-like solutions.

Thus, if the initial state is specified by a one-particle distribution function on allowed
configurations (\ref{Bog2_haos}), then the evolution of many hard spheres governed by the dual BBGKY
hierarchy (\ref{dh}) for marginal observables can be completely described by the generalized Enskog
kinetic equation (\ref{gke1}) and by the sequence of marginal functionals of state (\ref{f}).

\textcolor{blue!50!black}{\subsection{On the Boltzmann--Grad asymptotic behavior of the generalized Enskog equation}}
Let for initial one-particle distribution function (\ref{h2}) the assumption (\ref{lh2}) is valid.
Then for finite time interval the Boltzmann--Grad limit of dimensionless solution (\ref{F(t)1})
of the Cauchy problem of the non-Markovian Enskog kinetic equation (\ref{gke1}),(\ref{gkei})
exists in the same sense, namely
$\mathrm{w-}\lim_{\epsilon\rightarrow 0}\big(\epsilon^{2}F_{1}(t,x_1)-f_{1}(t,x_1)\big)=0,$
where the limit one-particle distribution function $f_{1}(t)$ is a weak solution of the Cauchy
problem of the Boltzmann kinetic equation (\ref{Bolz}),(\ref{Bolzi}).

As noted above, the all possible correlations of a system of hard spheres with inelastic collisions
are described by marginal functionals of the state (\ref{f}). Taking into consideration the fact of
the existence of the Boltzmann--Grad scaling limit of a solution of the non-Markovian Enskog kinetic
equation (\ref{gke1}), for marginal functionals of the state (\ref{f}) the following statement holds:
\begin{eqnarray*}
   &&\mathrm{w-}\lim\limits_{\epsilon\rightarrow 0}\big(\epsilon^{2s}
      F_{s}\big(t,x_1,\ldots,x_s\mid F_{1}(t)\big)-\prod\limits_{j=1}^{s}f_{1}(t,x_j)\big)=0,
\end{eqnarray*}
where the limit one-particle distribution function $f_{1}(t)$ is governed by the Boltzmann kinetic
equation with hard sphere collisions (\ref{Bolz}). This property of marginal functionals of state
(\ref{f}) means the propagation of the initial chaos \cite{CGP97}.

The proof of these statements is based on the properties of cumulants of asymptotically perturbed
groups of operators (\ref{nLkymyl}) and the explicit structure (\ref{skrrn}) of the generating
operators of series expansions (\ref{f}) for marginal functionals of state and of series (\ref{F(t)1}).


\bigskip
\textcolor{blue!50!black}{\section{Asymptotic behavior of hard spheres with inelastic collisions}}

In modern works \cite{BCP}, it is assumed that the microscopic dynamics of granular media is dissipative
and it can be described by a system of many hard spheres with inelastic collisions.

As is known \cite{CGP97}, in the Boltzmann-–Grad scaling limit the evolution of a one-dimensional
system of hard spheres with elastic collisions is trivial (free molecular motion or the Knudsen flow),
but in the case of inelastically collisions, the collective behavior of many hard rods is governed by
the Boltzmann kinetic equation with inelastic collisions. It should be emphasized that a system of many
hard rods with inelastic collisions displays the basic properties of granular gases. Below the approach
to the rigorous derivation of Boltzmann-type equation for one-dimensional granular gases will be outlined.

In the case of a one-dimensional granular gas for $t\geq0$ in dimensionless form the Cauchy problem
(\ref{gke1}),(\ref{gkei}) takes the form \cite{BG}:
\begin{eqnarray}\label{GE1}
  &&\hskip-8mm\frac{\partial}{\partial t}F_{1}(t,q_1,p_1)=
      -p_1\frac{\partial}{\partial q_1}F_{1}(t,q_1,p_1)+\\
  &&\int_0^\infty dP\,P\big(\frac{1}{(1-2\varepsilon)^2}\,
      F_{2}(t,q_1,p_1^\diamond(p_1,P),q_1-\epsilon,p_{2}^\diamond(p_1,P)\mid F_{1}(t))-\nonumber\\
  &&F_{2}(t,q_1,p_1,q_1-\epsilon,p_1+P\mid F_{1}(t))\big)+\nonumber
\end{eqnarray}
\begin{eqnarray}
  &&\int_0^\infty dP\,P \big(\frac{1}{(1-2\varepsilon)^2}\,F_{2}(t,q_1,\tilde p_1^\diamond(p_1,P),q_1+
      \epsilon,\tilde p_{2}^\diamond(p_1,P)\mid F_{1}(t))-\nonumber\\
  &&F_{2}(t,q_1,p_1,q_1+\epsilon,p_1-P\mid F_{1}(t))\big),\nonumber\\ \nonumber\\
\label{GKEi}
   &&F_{1}(t)|_{t=0}= F_{1}^{\epsilon,0},
\end{eqnarray}
where $\varepsilon=\frac{1-e}{2}\in [0,\frac{1}{2})$ and $e\in(0,1]$ is a restitution
coefficient, $\epsilon>0$ is a scaling parameter (the ratio of a hard sphere diameter (the length)
$\sigma>0$ to the mean free path), the collision integral is determined by marginal functional (\ref{f})
of the state $F_{1}(t)$ in the case of $s=2$ and the expressions:
\begin{eqnarray*}
   &&p_{1}^\diamond(p_1,P)=p_1-P+\frac{\varepsilon}{2\varepsilon -1}\,P,\\
   &&p_{2}^\diamond(p_1,P)=p_1-\frac{\varepsilon}{2\varepsilon -1}\,P\nonumber
\end{eqnarray*}
and the expressions
\begin{eqnarray*}
   &&\tilde p_{1}^\diamond(p_1,P)=p_1+P-\frac{\varepsilon}{2\varepsilon -1}\,P,\\
   &&\tilde p_{2}^\diamond(p_1,P)=p_1+\frac{\varepsilon}{2\varepsilon -1}\,P,\nonumber
\end{eqnarray*}
are transformed pre-collision momenta of inelastically colliding particles in a one-dimensional space.

If the initial one-particle distribution function satisfy the following condition:
$|F_{1}^{\epsilon,0}(x_1)|\leq Ce^{\textstyle-\frac{\beta}{2}{p^{2}_1}},$
where $\textstyle{\beta}>0$ is a parameter, $C<\infty$ is some constant, then every term of
the series
\begin{eqnarray}\label{ske1}
  &&\hskip-7mm F_{1}^{\epsilon}(t,x_1)=\sum\limits_{n=0}^{\infty}\frac{1}{n!}
     \int_{(\mathbb{R}\times\mathbb{R})^n}dx_2\ldots dx_{n+1}\,\mathfrak{A}_{1+n}^\ast(t)
     \prod_{i=1}^{n+1}F_{1}^{\epsilon,0}(x_i)\mathcal{X}_{\mathbb{R}^{(1+n)}\setminus \mathbb{W}_{1+n}},
\end{eqnarray}
exists, for finite time interval function (\ref{ske1}) is the uniformly convergent series with
respect to $x_1$ from arbitrary compact, and it is determined a weak solution of the Cauchy problem
of the non-Markovian Enskog equation (\ref{GE1}),(\ref{GKEi}).
We assume that in the sense of a weak convergence there exists the following limit:
\begin{eqnarray*}\label{asumdin}
   &&\mathrm{w-}\lim\limits_{\epsilon\rightarrow 0}\big(F_{1}^{\epsilon,0}(x_1)-f_{1}^0(x_1)\big)=0,
\end{eqnarray*}
then for finite time interval the Boltzmann--Grad limit of solution (\ref{ske1}) of the Cauchy problem
of the non-Markovian Enskog equation for a one-dimensional granular gas (\ref{GE1}) exists in the sense
of a weak convergence
\begin{eqnarray}\label{asymt}
   &&\mathrm{w-}\lim\limits_{\epsilon\rightarrow 0}\big(F_{1}^{\epsilon}(t,x_1)-f_{1}(t,x_1)\big)=0,
\end{eqnarray}
where the limit one-particle distribution function is determined by the uniformly convergent
on arbitrary compact set series:
\begin{eqnarray}\label{viter}
  &&\hskip-7mm f_{1}(t,x_1)=\sum\limits_{n=0}^{\infty}\frac{1}{n!}
     \int_{(\mathbb{R}\times\mathbb{R})^n}dx_2\ldots dx_{n+1}\,\mathfrak{A}_{1+n}^0(t)
     \prod_{i=1}^{n+1}f_{1}^0(x_i),
\end{eqnarray}
and the generating operator $\mathfrak{A}_{1+n}^0(t)\equiv\mathfrak{A}_{1+n}^0(t,1,\ldots,n+1)$
is the $(n+1)th$-order cumulant of adjoint semigroups $S_n^{\ast,0}(t)$ of point particles with
inelastic collisions. An infinitesimal generator of the semigroup of operators $S_n^{\ast,0}(t)$
is defined as the operator:
\begin{eqnarray*}
    &&\hskip-12mm(\mathcal{L}_{n}^{\ast,0}f_{n})(x_{1},\ldots,x_{n})=
       -\sum_{j=1}^{n}p_j\,\frac{\partial}{\partial q_j} f_{n}(x_{1},\ldots,x_{n})+\\
    &&\hskip-12mm\sum_{j_{1}<j_{2}=1}^{n}|p_{j_{2}}-p_{j_{1}}|\big(\frac{1}{(1-2\varepsilon)^{2}}
       f_{n}(x_{1},\ldots,x^{\diamond}_{j_{1}},\ldots,x^{\diamond}_{j_{2}},\ldots,x_{n})-
       f_{n}(x_{1},\ldots,x_{n})\big)\delta(q_{j_{1}}-q_{j_{2}}),
\end{eqnarray*}
where $x_j^\diamond\equiv(q_{j},p_{j}^\diamond)$ and the pre-collision momenta
$p^\diamond_{j_{1}},\,p^\diamond_{j_{2}}$ of inelastically colliding particles are determined by
the following expressions:
\begin{eqnarray*}
  &&p^{\diamond}_{j_{1}}=p_{j_{2}}+\frac{\varepsilon}{2\varepsilon -1}\,(p_{j_{1}}-p_{j_{2}}),\\
  &&p_{j_{2}}^{\diamond}=p_{j_{1}}-\frac{\varepsilon}{2\varepsilon-1}\,(p_{j_{1}}-p_{j_{2}}).\nonumber
\end{eqnarray*}

For $t\geq 0$ the limit one-particle distribution function represented by series (\ref{viter})
is a weak solution of the Cauchy problem of the Boltzmann-type kinetic equation of point particles
with inelastic collisions \cite{BG}
\begin{eqnarray}\label{Bolz1}
  &&\hskip-12mm\frac{\partial}{\partial t}f_1(t,q,p)=-p\,\frac{\partial}{\partial q}f_1(t,q,p)+
     \int_{-\infty}^{+\infty}d p_1\,|p-p_1|\times\\
  &&\hskip-12mm\big(\frac{1}{(1-2\varepsilon)^2}\,f_1(t,q,p^\diamond)\,f_1(t,q,p^\diamond_1)
     -f_1(t,q,p)\,f_1(t,q,p_1)\big)+\sum_{n=1}^{\infty}\mathcal{I}^{(n)}_{0}.\nonumber
\end{eqnarray}
In kinetic equation (\ref{Bolz1}) the remainder $\sum_{n=1}^{\infty}\mathcal{I}^{(n)}_{0}$
of the collision integral is determined by the expressions
\begin{eqnarray*}
  &&\hskip-12mm\mathcal{I}^{(n)}_{0}\equiv\frac{1}{n!}\int_0^\infty dP\,P\,
      \int_{(\mathbb{R}\times\mathbb{R})^{n}}dq_3dp_3\ldots dq_{n+2}dp_{n+2}\mathfrak{V}_{1+n}(t)\big(\frac{1}{(1-2\varepsilon)^2}
      f_1(t,q,p_1^\diamond(p,P))\times\nonumber\\
  &&\hskip-5mm f_1(t,q,p_{2}^\diamond(p,P))-f_1(t,q,p)f_1(t,q,p+P)\big)
      \prod_{i=3}^{n+2}f_{1}(t,q_i,p_i)+\nonumber\\
  &&\hskip-5mm\int_0^\infty dP\,P\,\int_{(\mathbb{R}\times\mathbb{R})^{n}}dq_3dp_3\ldots dq_{n+2}dp_{n+2}
      \mathfrak{V}_{1+n}(t)\big(\frac{1}{(1-2\varepsilon)^2}
      f_1(t,q,\tilde p_{1}^\diamond(p,P))\times\nonumber\\
  &&\hskip-5mm f_1(t,q,\tilde p_{2}^\diamond(p,P))-f_1(t,q,p)f_1(t,q,p-P)\big)
      \prod_{i=3}^{n+2}F_{1}(t,q_i,p_i),\nonumber
\end{eqnarray*}
where the generating operators $\mathfrak{V}_{1+n}(t)\equiv\mathfrak{V}_{1+n}(t,\{1,2\},3,\ldots,n+2),\,n\geq0,$
are represented by expansions (\ref{skrrn}) with respect to the cumulants of semigroups of scattering
operators of point hard rods with inelastic collisions in a one-dimensional space
\begin{eqnarray*}\label{scat}
   &&\widehat{S}_{n}^{0}(t,1,\ldots,n)\doteq
       S_n^{\ast,0}(t,1,\ldots,s)\prod_{i=1}^{n}S_1^{\ast,0}(t,i)^{-1}.
\end{eqnarray*}

In fact, the series expansions for the collision integral of the non-Markovian Enskog equation
for a granular gas or solution (\ref{ske1}) are represented as the power series over the density
so that the terms $\mathcal{I}^{(n)}_{0},\,n\geq1,$ of the collision integral in kinetic equation
(\ref{Bolz}) are corrections with respect to the density to the Boltzmann collision integral for
one-dimensional granular gases stated in \cite{BCP},\cite{T04}.

Since the scattering operator of point hard rods is an identity operator in the aproximation of
elastic collisions, namely, in the limit $\varepsilon\rightarrow0$, the collision integral of the
Boltzmann kinetic equation (\ref{Bolz1}) in a one-dimensional space is identical to zero. In the
quasi-elastic limit \cite{T04} the limit one-particle distribution function (\ref{viter})
\begin{eqnarray*}
   &&\lim_{\varepsilon \rightarrow 0}\varepsilon f_1(t,q,v)=f^0(t,q,v),
\end{eqnarray*}
satisfies the nonlinear friction kinetic equation for granular gases of the following form \cite{T04}
\begin{eqnarray*}
   &&\hskip-5mm\frac{\partial}{\partial t}f^0(t,q,v)=-v\,\frac{\partial}{\partial q} f^0(t,q,v) +
       \frac{\partial}{\partial v}\int_{-\infty}^\infty dv_1\,
       |v_1-v|\,(v_1-v)\, f^0(t,q,v_1)f^0(t,q,v).
\end{eqnarray*}

Taking into consideration result (\ref{asymt}) on the Boltzmann--Grad asymptotic behavior
of the non-Markovian Enskog equation (\ref{gke1}), for marginal functionals of state
(\ref{f}) in a one-dimensional space the following statement is true \cite{BG}:
\begin{eqnarray}\label{lf}
  &&\hskip-9mm\mathrm{w-}\lim\limits_{\epsilon\rightarrow 0}
      \big(F_{s}\big(t,x_1,\ldots,x_s\mid F_{1}^{\epsilon}(t)\big)
      -f_{s}\big(t,x_1,\ldots,x_s\mid f_{1}(t)\big)\big)=0,\quad s\geq2,
\end{eqnarray}
where the limit marginal functionals (\ref{lf}) with respect to limit one-particle distribution
function (\ref{viter}) are determined by the series expansions with the structure similar to
series (\ref{f}) and the generating operators represented by expansions (\ref{skrrn}) over the
cumulants of semigroups of scattering operators of point hard rods with inelastic collisions
in a one-dimensional space.

Note that, as mention above, in the case of a system of hard rods with elastic collisions
the limit marginal functionals of the state are the product of the limit one-particle
distribution functions, describing the free motion of point particles.

Thus, the Boltzmann--Grad asymptotic behavior of solution (\ref{ske1}) of the non-Markovian
Enskog equation (\ref{GE1}) is governed by the Boltzmann kinetic equation (\ref{Bolz1}) for
a one-dimensional granular gas.

We emphasize that the Boltzmann-type equation (\ref{Bolz1}) describes the memory effects
in a one-dimensional granular gas. In addition, the limit marginal functionals of state
$f_{s}\big(t,x_1,\ldots,x_s\mid f_{1}(t)\big),\,s\geq2,$ which are defined above, describe
the process of the propagation of initial chaos in a one-dimensional granular gase, or,
in other words, the process of creation correlations in a system of hard rods with
inelastic collisions.

It should note that the Boltzmann--Grad asymptotic behavior of the non-Markovian Enskog
equation with inelastic collisions in a multidimensional space is analogous of the Boltzmann--Grad
asymptotic behavior of a hard sphere system with the elastic collisions \cite{BG}, i.e. it is
governed by the Boltzmann equation for a granular gas \cite{V06}, and the asymptotics of marginal
functionals of state (\ref{f}) is the product of its solution.

\newpage
\textcolor{blue!50!black}{\section{Outlook}}

In the paper, two new approaches to the description of the kinetic evolution of
many-particle systems with hard sphere collisions were outlooked. One of them is
a formalism for the description of the evolution of infinitely many hard spheres
within the framework of marginal observables in the Boltzmann--Grad scaling limit.
Another approach to the description of the kinetic evolution of hard spheres is
based on the non-Markovian generalization of the Enskog kinetic equation.

In particular, it was established that a chaos property of the Boltzmann--Grad scaling
behavior of the $s$-particle marginal distribution function of infinitely many hard
spheres is equivalent in the sense of equality (\ref{dchaos}) to a solution of the dual
Boltzmann hierarchy (\ref{vdh}) in the case of the $s$-ary marginal observable. In the
case of additive-type marginal observables, the evolution is equivalent to the evolution
of the state governed by the Boltzmann equation.
We remark that a similar approach to the description of quantum many-particle systems
in a mean-field limit was developed in the paper \cite{G11} (see also \cite{GerUJP}).

One of the advantages of the considered approaches is the possibility to construct
the kinetic equations in scaling limits, involving correlations at the initial time,
which can characterize the condensed states of a system of colliding particles.

We emphasize that the approach to the derivation of the Boltzmann equation from
underlying dynamics governed by the generalized Enskog kinetic equation enables
us to construct also the higher-order corrections to the Boltzmann--Grad evolution
of many-particle systems with hard sphere collisions.

Some applications of the discussed methods to the derivation of kinetic equations
for different nature of many-particle systems with collisions are considering
in the papers \cite{GF},\cite{GG12}.


\bigskip

\textcolor{blue!50!black}{\addcontentsline{toc}{section}{References}}
\renewcommand{\refname}{\textcolor{blue!50!black}{References}}

\end{document}